\documentclass[usenatbib]{mn2e}
\usepackage{amsmath}
\usepackage{graphicx}
\usepackage{epsfig}
\newcommand{\Vmax}{$V_{\mathrm{max}}$\,}

\title[Groups of two galaxies in SDSS: implications of colours on star formation quenching]{Groups of two galaxies in SDSS: implications of colours on star formation quenching time-scales}
\author[Trinh et al.]{Christopher Q. Trinh$^{1}$\thanks{E-mail:
c.trinh@physics.usyd.edu.au}, 
Elizabeth J. Barton$^{2}$, James S. Bullock$^{2}$, Michael C. Cooper$^{2}$,
\newauthor Andrew R. Zentner$^{3}$ and Risa H. Wechsler$^{4}$ \\
$^{1}$Sydney Institute for Astronomy, School of Physics, The University of Sydney, NSW 2006, Australia\\
$^{2}$Center for Cosmology, Department of Physics and Astronomy, University of California, Irvine, CA 92697-4575\\
$^{3}$Department of Physics and Astronomy, University of Pittsburgh, Pittsburgh, PA 15260\\
$^{4}$Kavli Institute for Particle and Astrophysics and Cosmology, Physics Department, Stanford University, Stanford, CA 94305}
\begin{document}

\date{2013 January 24}

\pagerange{\pageref{firstpage}--\pageref{lastpage}} \pubyear{2013}

\maketitle

\label{firstpage}

\begin{abstract}
We have devised a method to select galaxies that are isolated in their dark matter halo ($N=1$ systems) and galaxies that reside in a group of exactly two ($N=2$ systems). Our $N=2$ systems are widely-separated (up to $\sim$\,200\,$h^{-1}$\,kpc), where close galaxy-galaxy interactions are not dominant. We apply our selection criteria to two volume-limited samples of galaxies from SDSS DR6 with $M_{r}-5 \log_{10} h \leq$ $-19$ and $-20$ to study the effects of the environment of very sparse groups on galaxy colour. For satellite galaxies in a group of two, we find a red excess attributed to star formation quenching of 0.15\,$\pm$\,0.01 and 0.14\,$\pm$\,0.01 for the $-19$ and $-20$ samples, respectively, relative to isolated galaxies of the same stellar mass. Assuming $N=1$ systems are the progenitors of $N=2$ systems, an immediate-rapid star formation quenching scenario is inconsistent with these observations. A delayed-then-rapid star formation quenching scenario with a delay time of 3.3 and 3.7\,Gyr for the $-19$ and $-20$ samples, respectively, yields a red excess prediction in agreement with the observations. The observations also reveal that central galaxies in a group of two have a slight blue excess of 0.06\,$\pm$\,0.02 and  0.02\,$\pm$\,0.01 for the $-19$ and $-20$ samples, respectively, relative to $N=1$ populations of the same stellar mass. Our results demonstrate that even the environment of very sparse groups of luminous galaxies influence galaxy evolution and in-depth studies of these simple systems are an essential step towards understanding galaxy evolution in general. 
\end{abstract}


\section{Introduction}\label{sec:intro}
The overall galaxy population out to $z\sim1$ divides into two distinct types: early-type and late-type galaxies \citep[e.g.][]{hubble1926,strateva2001,blanton2003,bell2004,tanaka2005,baldry2006,cooper2006}. Although the exact definition of the two types varies in the literature, late-type galaxies tend to have spiral morphology and are blue due to a highly active recent star formation history. On the other hand, early-type galaxies tend to have elliptical morphology and are red due to a lack of recent star formation. 


From observations of galaxies at different redshifts we know that the stellar mass on the red sequence has increased by a factor of 2 since $z\sim 1$ \citep{bell2004}. Along with numerous other observations this suggests that physical processes have been at work since $z\sim 1$ transforming late-type galaxies into early-type galaxies. Young clusters at $z\sim0.5$ contain many spiral galaxies and very few S0 galaxies while local clusters have a ratio of S0 to elliptical galaxies greater by a factor of 5 and a similar decrease in the number of spiral galaxies \citep{dressler1997}. Also, the galaxy luminosity function density normalization for blue galaxies has remained essentially constant, while it has doubled ($\sim0.5$\,dex) for red galaxies since $z\sim1$ \citep{faber2007}. 




In the currently favoured $\Lambda$ Cold Dark Matter ($\Lambda$CDM) picture of structure formation \citep{blumenthal1984}, dark matter collapses into haloes, trapping baryonic matter at early epochs. In galaxy-sized dark matter haloes, cool baryonic material falls in while retaining angular momentum to form luminous spiral galaxies. Galaxy groups and cluster form hierarchically when smaller dark matter haloes combine to form larger dark matter haloes \citep{white1978,kauffmann1993,cole2000}. This picture motivates a number of physical processes that may alter a galaxy's morphology and/or quench star formation through the removal and/or depletion of the galaxy's gas reservoirs bringing about a late-type to early-type transition. 


The merger of two dark matter haloes will eventually result in the merger of their central galaxies as a result of dynamical friction \citep{chandrasekhar1943}, potentially permanently altering both morphology and star formation rates. Even before the central galaxies merge they are altered by interactions, especially if a smaller dark matter halo has fallen into a large group or cluster-sized halo with a dense intracluster medium and many other un-merged satellite galaxies. These environmental interactions broadly fall into two categories: gravitational tides and gas collisions. 

Tidal interactions due to the main cluster potential on the infalling satellite galaxy may strip it of stars and/or gas leading to changes in morphology and/or star formation rates \citep{fujita1998}. Tidal interactions from a close galaxy pass also triggers inflows of gas into the centres of galaxies, inducing short-lived starburst \citep{mihos1996}. Once the galaxy has exhausted its supply of gas, it will quickly redden unless it can replenish its supply of gas. One mechanism that can prevent the accretion of new gas is feedback from an active galactic nucleus \citep[AGN, e.g.][]{dimatteo2005,bower2006,croton2006}. The energy released by an AGN generates an outflow of gas preventing accretion, which may transform a blue, active galaxy into a red and dead galaxy although the details remain unclear. In addition to tidal interactions with the main cluster potential, a satellite galaxy will tidally interact with the other un-merged satellites within the halo. All these close high-speed encounters, which occur approximately once per Gyr, are referred to as galaxy harassment \citep[e.g][]{farouki1981,moore1996} and can transform small disc galaxies into dwarf elliptical or dwarf spheroidal galaxies.  

The collisional interaction of the hot gaseous intracluster medium of the larger host halo with the cold disc gas and/or the hot halo gas reservoir of the smaller halo's central galaxy may rapidly strip it of either gas reservoir in a process known as ram-pressure stripping \citep{gunn1972}. High-resolution hydrodynamical simulations have shown that ram-pressure stripping can remove the entire
\textsc{H\,I} gas content within 100\,Myr from a luminous spiral galaxy like the Milky Way leading to redder colours from a lack of star formation \citep{quilis2000}. However, the morphology of the original stellar disc is unaffected by ram-pressure stripping, which results in the spiral galaxy being transformed into an S0-like galaxy. If only the hot halo gas reservoir of a satellite galaxy, which replenishes the cold disc gas that is converted into stars, is removed by ram-pressure stripping or tides, the result is a gradual decline of star formation because it continues until the cold disc gas is exhausted. To distinguish it from the removal of cold disc gas, which results in a rapid truncation in star formation, this is usually referred to as strangulation \citep{larson1980,balogh2000}. Strangulation should result in a similar transformation of a spiral into an S0-like galaxy, except on a different time-scale than ram-pressure stripping of cold disc gas. 

If an infalling satellite galaxy is able to withstand the environmental interactions discussed above, a major or minor merger with another satellite galaxy \citep{makino1997} or the central galaxy of the host halo may result in a merger remnant with drastically different morphology and/or star formation rate than either of the two progenitors. The severity of the change is highly dependent on the mass ratio of the two progenitors. If a spiral galaxy is involved in a minor merger, a temporary starburst may be triggered from the compression of colliding cold gas and the disc morphology may be affected. However, morphology is definitely altered when two spiral galaxies with similar masses experience a major merger. In this case, the merger remnant may have spheroidal or elliptical morphology \citep{toomre1972}. If the gas reservoirs depleted in the triggered starburst are not replenished because of say AGN feedback, the major merger will have transformed two active spiral galaxies into a red and dead spheroidal or elliptical galaxy. 

A fundamental goal of astrophysics is a comprehensive understanding of the role mergers and environmental processes play in building and evolving the diverse set of galaxies that exist in the Universe. Studies of galaxy evolution span a range of environments including isolated galaxies \citep[e.g.][]{allam2005,tollerud2011,edman2012}, close pairs \citep[e.g.][]{solalonso2006,barton2007,ellison2008,ellison2010,ellison2011,patton2011,scudder2012}, groups \citep[e.g.][]{balogh2004a,gerke2005,weinmann2006a,weinmann2006b,weinmann2009,kang2008,vdb2008,kimm2009,skibba2009,iovino2010,pasquali2010,mcgee2011,wetzel2012,knobel2013} and clusters \citep[e.g.][]{balogh2002,rines2005,tanaka2005,vonderlinden2007,vonderlinden2010}. 

However, a general understanding of galaxy evolution requires an understanding of the simplest galactic environments, which are isolated galaxies residing alone in their dark matter halo ($N=1$ system) and systems of two galaxies sharing a dark matter halo ($N=2$ system), which includes galaxy pairs and groups of two galaxies. Isolated galaxies are the least ambiguous and are the most controlled environments in which to study galaxy evolution. In general, more isolated galaxies are more likely to have later-type morphologies, higher star formation rates and bluer colours \citep[e.g.][]{postman1984,blanton2005b}. 

In this paper, we present an preliminary study of the difference between the properties of isolated galaxies and groups of two galaxies in order to study the effects of the environment of very sparse groups. We use a cosmological model consisting of a hybrid $N$-body/semi-analytic substructure simulation to understand and remove the contamination from galaxies in other environments allowing us to study the full, uncontaminated distributions of star-forming and morphological parameters instead of just the average trend. This approach differs from previous studies that utilise group-finding algorithms \citep[e.g.][]{yang2005}, which have difficulty accounting for interloper systems. Our technique has been used previously by \citet{barton2007} to study triggered star formation in close galaxy pairs, where galaxy-galaxy interactions are dominant. In galaxy groups, where galaxy-galaxy interactions are not dominant, the dominant mechanism responsible for star formation quenching is still an open question and a very active field of research \citep[e.g.][]{weinmann2006a,weinmann2006b,weinmann2009,kang2008,vdb2008,kimm2009,skibba2009,pasquali2010,mcgee2009,mcgee2011,wetzel2012,phillips2013}. The low-mass haloes of groups of two galaxies examined in this paper represent the minimum mass where the dominant group mechanism begins to activate \citep{mcgee2009}. Investigations of groups of two aim to shed light on the subject and they afford a number of other advantages including the fact that $\Lambda$CDM makes robust predictions for the merger histories of $N=2$ systems \citep[e.g.][]{stewart2008} and such systems are routinely studied in high-resolution hydrodynamical simulations \citep[e.g.][]{cox2006} allowing for a direct comparison yielding insights into our assumptions and treatment of interactions, star formation and feedback. As such, groups of two galaxies are a valuable vehicle for the study of galaxy evolution and will contribute toward a comprehensive understanding of the topic in general.  

The layout of this paper is as follows. In Section \ref{sec:cosmological} we discuss the cosmological model we use throughout the paper to yield information on several important unobservable galaxy properties. Section \ref{sec:selection} describes the selection of populations of isolated galaxies and groups of two galaxies starting with an analysis of the environment of the simulated galaxies in our cosmological model. Section \ref{sec:sdss} contains a description of the observational data set resulting from the application of our selection criteria to galaxies from the Sloan Digital Sky Survey \citep[SDSS,][]{york2000}. Our method for correcting for the contamination by galaxies in other environments is explained in Section \ref{sec:contamination}.  We look at the stellar mass distribution of isolated galaxies and groups of two galaxies and discuss our Monte Carlo mass resampling technique for producing populations of the same stellar mass in Section \ref{sec:stellar}. Results are discussed in Section \ref{sec:results}. In Section \ref{sec:colour} we concern ourselves with the difference in colour between isolated galaxies and groups of two galaxies. In Section \ref{sec:satellites} we discuss the red excess of satellite galaxies in a group of two and its implications on star formation quenching time-scales. The blue excess of central galaxies in a group of two and its origins are explored in Section \ref{sec:centrals} followed by a conclusion in Section \ref{sec:conclusion}. 

\section{Cosmological model}\label{sec:cosmological}



The main components of our cosmological model include a large $N$-body simulation of cold dark matter halo formation and a semi-analytic cold dark matter substructure model \citep{zentner2003,zentner2005}. This model has been previously discussed in \citet{berrier2006} and \citet{barton2007} and we provide a brief review here.

The model uses $N$-body simulations to characterize the spatial and mass distributions of ``host'' dark matter haloes. By definition, the centres of host haloes do not lie within the virial radius of other haloes. Our $N$-body simulation follows the evolution of 512$^{3}$ dark matter particles until $z=0$ in a comoving box of volume 120\,$h^{-1}$\,Mpc$^{3}$ using the Adaptive Refinement Tree code of \citet{kravtsov1997} in a standard $\Lambda$CDM cosmology with $\Omega_{m}=0.3$, $\Omega_{\Lambda}=0.7$, $h=0.7$, and $\sigma_{8}=0.9$. The implied particle mass is $m_{p} \simeq 1.1\times 10^{9}$\,$h^{-1}$\,$M_{\odot}$ and the simulation grid is refined down to a minimum cell size of $h_{\textrm{peak}}\simeq 1.8$\,$h^{-1}$\,kpc on a side. For more details regarding the $N$-body simulation see \citet{zentner2005}, \citet{allgood2006} and \citet{wechsler2006}.

$N$-body simulations predict the substructure content of host haloes, but suffer from various numerical resolution limitations. Thus, we populate each host dark matter halo in the simulation volume with subhaloes using the semi-analytic substructure model of \citet{zentner2005}, which effectively has infinite resolution. We use the semi-analytic model to randomly generate four independently realized mass accretion histories using the stochastic method of \citet{somerville1999} for each host halo of mass $M$ at redshift $z$ in our simulation volume. Then, we determine the orbital evolution for each infalling subhalo in the potential of the host from accretion until $z=0$, during which it loses mass and its maximum circular velocity decreases as its profile is heated by interactions. If the maximum circular velocity of a subhalo falls below 60\,km\,s$^{-1}$ at any time, it is removed to mimic the dissolution of the observable galaxy as a result of these interactions. Each mass accretion history results in a different subhalo population in the simulation volume. We interpret these realizations as four independent cosmological volumes with identical large-scale structure, but different small-scale structure. All four realizations will be used in the statistical analysis that follows. 

\begin{figure*}
\center
\begin{tabular}{cc}
\includegraphics[width=0.45\textwidth]{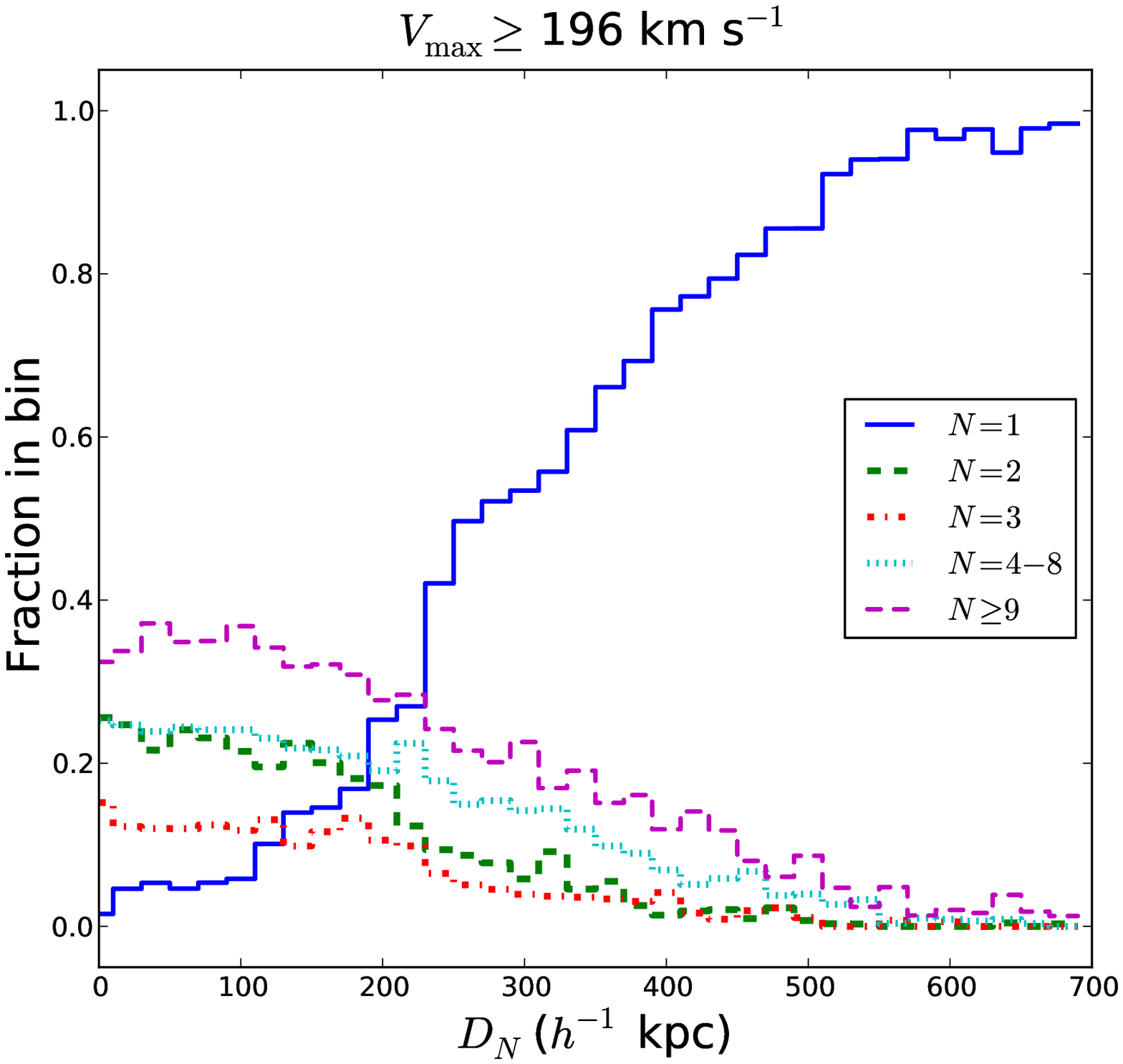}
\includegraphics[width=0.45\textwidth]{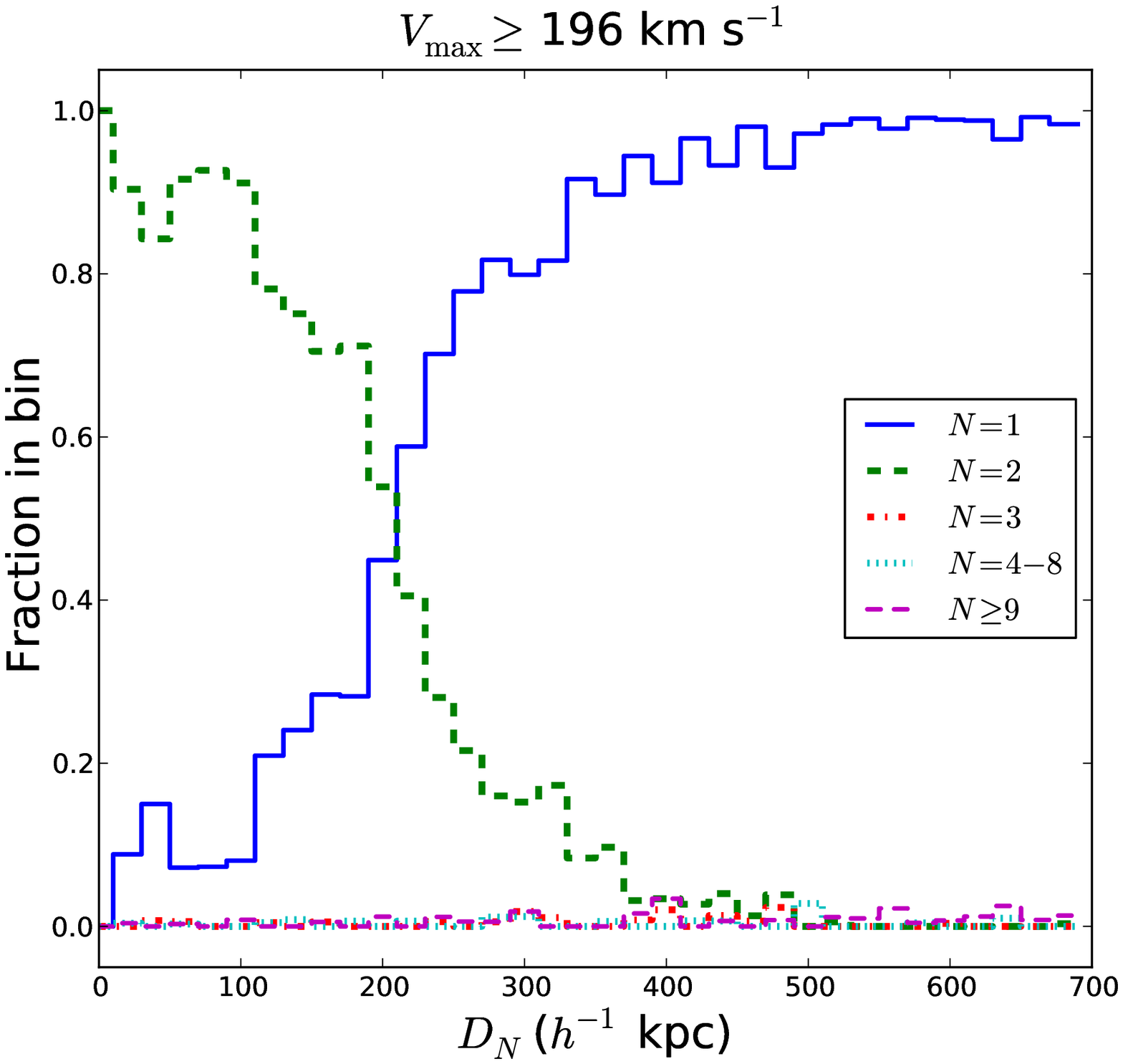}
\end{tabular}
\caption{(Left) Fraction of simulated galaxies in our mock volume-limited redshift survey ($M_{r}-5 \log_{10} h \leq -20$ or \Vmax$\geq196$\,km\,s$^{-1}$) that reside in a dark matter halo of a given multiplicity $N$, as a function of the comoving projected distance to their nearest neighbour, $D_{N}$. Each bin in $D_{N}$ is normalised to 1. $D_{N}$ alone cannot be used to define samples of galaxies in $N=2$ systems from volume-limited redshift surveys. Galaxies with small $D_{N}$ values are predominantly in haloes with a total of 9 or more galaxies. (Right) Fraction of simulated galaxies with exactly one neighbour within 700\,$h^{-1}$\,kpc ($N_{700}=1$) that reside in a dark matter halo of a given multiplicity $N$, as a function of $D_{N}$. Very few galaxies with $N_{700}=1$ reside in haloes with more than two galaxies. Galaxies with $N_{700}=1$ and large values of $D_{N}$ have a very high probability of being isolated in their halo and those with $N_{700}=1$ and small values of $D_{N}$ have a high probability of being in an isolated galaxy pair.\label{Dn_dist}}
\end{figure*}

We assume each subhalo in the simulation volume contains a luminous satellite galaxy and each host halo has a luminous central galaxy. For host haloes, we use the maximum circular velocity as a proxy for luminosity and assume a monotonic relationship between the two following previous works \citep[e.g.][]{conroy2006,berrier2006,barton2007,berrier2011,tollerud2011}. For subhaloes, we use the maximum circular velocity at accretion \citep[see][]{berrier2006}, which has been shown to reproduce the galaxy two-point correlation function at many epochs \citep{conroy2006}. In other words, all haloes above a given cutoff \Vmax will correspond to a population of galaxies brighter than some absolute magnitude. By exploring the number density of haloes in the model as a function of $V_{\mathrm{max}}$, we find that the number density of haloes with \Vmax $\geq$\,140\,(196)\,km\,s$^{-1}$ matches the observed number density of SDSS Data Release 6 \citep[SDSS DR6,][]{amc2008} volume-limited samples with $M_{r} - 5\log_{10} h \leq$\,$-19$\,($-20$). 

With luminous galaxies mapped onto dark matter haloes, we construct a mock volume-limited galaxy redshift survey to mimic galaxies from SDSS DR6 by placing the simulation volume the appropriate distance from the observer and computing the right ascension, declination and redshift of each simulated galaxy. This technique has been previously used to mimic galaxies from SDSS DR4 \citep{barton2007,berrier2011} and DR7 \citep{tollerud2011} including selection effects and we refer the reader to these works for more details. 

For each simulated galaxies in the mock volume-limited redshift survey, we are able to quantify the number of luminous galaxies that share the same host halo, $N$, which is an unobservable parameter in real redshift surveys that is a key measure of the environment of a galaxy. We analyse the correlation between $N$ and observable measures of environment, such as projected nearest neighbour distance. This analysis reveals selection criteria for defining relatively pure populations of isolated galaxies ($N=1$ systems) and groups of two galaxies ($N=2$ systems) from volume-limited redshift surveys and gives statistical information on other unobservables of these populations. 

Note that mock redshift surveys using similar hybrid $N$-body/semi-analytic substructure simulations have been used previously by \citet{barton2007} to study triggered star formation in close galaxy pairs and by \citet{tollerud2011} to study Milky Way/Large Magellanic Cloud-like systems. The work on groups of two galaxies here is a natural extension of these studies using similar methods. It is possible to construct mock redshift surveys using high-resolution large-box simulations \citep[e.g.][]{berrier2011,tollerud2011}, such as the Millennium-II simulations \citep{boylankolchin2009}, where semi-analytic models are not required because subhaloes are well-resolved and these will be consider in future studies of groups of two galaxies. 

\begin{table*}
\begin{minipage}{165mm}
\center
\caption{Sample definitions, sizes and purity. Symbol definitions are as follows: $D_{N}$ - comoving projected distance to the nearest neighbour, $N_{700}$ - number of neighbours within a comoving projected distance of 700\,$h^{-1}$\,kpc, $N_{\mathrm{SDSS}}$ - number of galaxies in the SDSS sample, $N_{\mathrm{sim}}$ - number of simulated galaxies in the cosmological model, $f_{N=1}$ - fraction of simulated galaxies in $N=1$ systems, $f_{N=2}$ - fraction of simulated galaxies in $N=2$ systems, $f_{N>2}$ - number of simulated galaxies in $N>2$ systems\label{sampletable}}
\begin{tabular}{@{}cccccccccc@{}}
\hline
Sample & Magnitude Limit & $D_{N}$ ($h^{-1}$\,kpc) & $N_{700}$ & $N_{\textrm{SDSS}}$ & $N_{\textrm{sim}}$ & $f_{N=1}$ & $f_{N=2}$ & $f_{N> 2}$\\
 \hline
$N=1$ & $\leq -20$ & $\geq 400$ & $\leq 1$ & 26,192 & 18,030 &  0.995 & 0.001 & 0.004 \\
$N=1$ & $\leq -19$ & $\geq 400$ & $\leq 1$ & 7,266 & 37,309 &  0.999 & 0.000 & 0.000 \\
$N=2$ & $\leq -20$ & $\leq 250$ & 1 & 3,229 & 2,675 &  0.203 & 0.788 & 0.008 \\
$N=2$ & $\leq -20$ & $\leq 200$ & 1 & 2,549 & 2,276 &  0.144 & 0.848 & 0.007 \\
$N=2$ & $\leq -20$ & $\leq 150$ & 1 & 1,830 & 1,807 &  0.106 & 0.887 & 0.007 \\
$N=2$ & $\leq -19$ & $\leq 250$ & 1 & 1,096 & 5,634 &  0.369 & 0.629 & 0.002 \\
$N=2$ & $\leq -19$ & $\leq 200$ & 1 & 918 & 4,786 &  0.289 & 0.709 & 0.002 \\
$N=2$ & $\leq -19$ & $\leq 150$ & 1 & 721 & 3,769 &  0.199 & 0.799 & 0.002 \\
\hline
\end{tabular}
\end{minipage}
\end{table*}

\section{Selection criteria for $N=1$ and $N=2$ systems}\label{sec:selection}
First, we analyse the environment of galaxies in our mock volume-limited redshift surveys in order to define selection criteria for isolated galaxies ($N=1$) and groups of two galaxies ($N=2$) in SDSS. Following \citet{barton2007}, for each ``galaxy'' (halo or subhalo with \Vmax $\geq$ 140 or 196\,km\,s$^{-1}$) in our mock volume-limited sample, we compute $D_{N}$, the comoving projected distance to the galaxy's nearest neighbour within $\Delta V \leq 1000$\,km\,s$^{-1}$, using periodic boundary conditions to fully sample its environment when necessary. We also compute $N_{700}$, the total number of galaxies within a comoving projected distance of 700\,$h^{-1}$\,kpc within $\Delta V \leq 1000$\,km\,s$^{-1}$. Lastly, we measure $N$, the total number of galaxies that lie within the same host dark matter halo, which is unobservable in galaxy redshift surveys. Constraints on the observables $D_{N}$ and $N_{700}$ can identify galaxies in redshift surveys with a high probability of being in an $N=1$ or $N=2$ system, by applying them simultaneously to the data and the simulations.

Following \citet{barton2007}, but with a different \Vmax cutoff, Figure \ref{Dn_dist} (left) shows the fraction of galaxies with \Vmax $\geq$ 196\,km\,s$^{-1}$ in our mock volume-limited sample that reside in a halo of a given multiplicity $N$ as a function of $D_{N}$. Galaxies that are relatively far from their nearest neighbour are overwhelmingly in $N=1$ systems. Galaxies that are relatively close to their nearest neighbour are predominantly in haloes with nine or more galaxies in the same dark matter halo. The results are qualitatively similar for \Vmax $\geq$ 140\,km\,s$^{-1}$. Thus, using $D_{N}$ \emph{alone} does not provide a sufficient means to select for $N=2$ systems in volume-limited redshift surveys. 

However, these systems can be selected using $D_{N}$ and $N_{700}$ together. Figure \ref{Dn_dist} (right) shows the fraction of galaxies in our mock volume-limited sample with $N_{700} = 1$ that reside in a halo of a given multiplicity $N$ as a function of $D_{N}$. The plot shows that very few galaxies with $N_{700}=1$ reside in haloes with $N> 2$. Moreover, galaxies with $N_{700}=1$ and large values of $D_{N}$ are in $N=1$ haloes and those with small values of $D_{N}$ reside almost exclusively in $N=2$ haloes. Thus, we identify a population of galaxies with a high probability of residing in an $N=2$ system can be defined  by selecting galaxies from volume-limited redshift surveys with exactly one neighbour within a projected distance of 700\,$h^{-1}$\,kpc and $D_{N}$ smaller than some maximum value. The purity and contamination by interlopers in the $N=2$ sample may be determined from our mock redshift survey by quantifying the number of galaxies residing in $N=1$, $N=2$, and $N>2$ haloes for a given maximum $D_{N}$. Table \ref{sampletable} lists the fraction of $N=1$ haloes, $f_{N=1}$, the fraction of $N=2$ haloes, $f_{N=2}$, and the fraction of $N>2$ haloes, $f_{N>2}$, for a maximum $D_{N}$ of 250, 200, and 150\,$h^{-1}$\,kpc for both luminosities. The $D_{N}\leq$ 200\,$h^{-1}$\,kpc population is our fiducial $N=2$ sample but we verify all results for the other two values.



A population of galaxies with a very high probability of being isolated in $N=1$ systems results from selecting galaxies with $N_{700} \leq 1$ and $D_{N} \geq$ 400\,$h^{-1}$\,kpc. The purity and contamination for this population for both luminosities are also listed in Table \ref{sampletable}. Note that the $N=1$ population defined in this way is very pure (99.5--99.9 per cent).  


\subsection{$N=1$ and $N=2$ systems in SDSS}\label{sec:sdss}
Our goal is to identify pure samples of $N=1$ and $N=2$ galaxies in order to investigate the differences in their properties. We apply the selection criteria discussed in the previous section to a volume-limited sample of galaxies from the New York University Value-Added Galaxy Catalogue \citep[NYU-VAGC,][]{blanton2005} based on SDSS DR6 with $M_{r} - 5\log_{10} h \leq$\,$-19$\,($-20$). The volume-limited samples contain galaxies from the Main galaxy sample with an extinction-corrected apparent magnitude $r \leq 17.77$ in regions of redshift completeness greater than 0.8. The volume-limited sample with $M_{r} - 5\log_{10} h \leq$\,$-19$\,($-20$) contains 67,472\,(107,327) galaxies.

First, we define a population of $N=1$ galaxies from the volume-limited sample. For each galaxy in the volume-limited sample, we identify all other galaxies within $\Delta V\leq 1000$\,km\,s$^{-1}$ and compute $D_{N}$ and $N_{700}$. We also compute the number of potential neighbours within a projected distance of 700\,$h^{-1}$\,kpc. Potential neighbours are galaxies in the NYU-VAGC without a measured redshift but in the relevant magnitude range, i.e. have a maximum $K$-corrected apparent magnitude of
\begin{equation}
r_{\mathrm{max}} = 5\log_{10}\left(\frac{d_{\mathrm{comoving}}}{1\,\mathrm{Mpc}}\right)+25+0.32+M_{r},
\end{equation}
where $d_{\mathrm{comoving}}$ is the comoving distance of the galaxy in the volume-limited survey in Mpc and $M_{r}$ is the maximum absolute magnitude of the volume-limited survey. 

Next, we apply our selection criteria on $D_{N}$ and $N_{700}$ as defined from our mock volume-limited redshift survey. However, the mock volume-limited redshift survey does not suffer from incompleteness as SDSS does, which affects the $N_{700}$ environment statistic as described in \citet{berrier2011}. To account for incompleteness, we follow \citet{berrier2011} and use four random catalogues of evenly distributed galaxies, provided by the NYU-VAGC website \footnote{http://sdss.physics.nyu.edu/vagc/}, to estimate the completeness of the 700\,$h^{-1}$\,kpc, $\Delta V=\pm1000$\,km\,s$^{-1}$ cylinder for each galaxy in the volume-limited sample. Each random galaxy is weighted by the completeness of the sector from the {\sc FGOTMAIN} parameter tabulated in the NYU-VAGC and by an estimate of the fraction of the luminosity function \citep{blanton2005} missed due to the limiting magnitude of the sector. Then, we sum the number of weighted random galaxies and normalise to the area searched on the sky. We use the random counts as a measure of the local completeness of the survey. In the following, we do not consider NYU-VAGC galaxies with a local completeness less than 1.5$\sigma$ the mean local completeness level.

A population of $N=1$ galaxies is defined by selecting all galaxies with $N_{700}\leq 1$ and $D_{N}\geq$ 400\,$h^{-1}$\,kpc and no potential neighbours. The SDSS $-19$\,($-20$) $N=1$ sample contains 7,266\,(26,192) galaxies. The expected purity of this population is $f_{N=1}=$ 0.999\,(0.995) with negligible amounts of contamination by $N=2$ and $N>2$ systems according to our mock volume-limited redshift survey (see Table \ref{sampletable}). 

A population of $N=2$ galaxies is defined by selecting all galaxies with $N_{700}=1$ and $D_{N}\leq$ 200\,$h^{-1}$\,kpc and no potential neighbours. The SDSS $-19$\,($-20$) $N=2$ sample contains 918\,(2,549) galaxies. The expected purity of this sample is $f_{N=2}=$ 0.709\,(0.848) and the contamination is mostly from $N=1$ systems, $f_{N=1}=$ 0.289\,(0.144), according to our mock redshift survey. The contamination by systems with $N>2$, $f_{N>2}=$ 0.002\,(0.007), is very small and we will only directly consider and correct the contamination by $N=1$ systems. If instead we take the maximum $D_{N}$ for the $N=2$ sample to be 250 or 150\,$h^{-1}$\,kpc, the sample then contains 1,096\,(3,229) and 721\,(1,830) galaxies, respectively. However, the contamination is higher for 250\,$h^{-1}$\,kpc with $f_{N=1}=$ 0.369\,(0.203) and lower for 150\,$h^{-1}$\,kpc with $f_{N=1}=$ 0.199\,(0.106). 

The overall $N=2$ population may be further divided into satellite and central galaxies. The satellite galaxy resides within a smaller subhalo, which has fallen into the larger host halo of the central galaxy. For each $N=2$ pair, the less\,(more) luminous galaxy in $M_{r}$ is selected as the satellite\,(central) galaxy. Note that in some cases, the two galaxies are both central galaxies in separate dark matter haloes or may be both satellites in a richer system. We have quantified the frequency of these occurrences in Table \ref{sampletable}. The contamination by $N=1$ systems, i.e. when the two galaxies are both centrals is statistically-corrected as described in Section \ref{sec:contamination}.

\begin{figure*}
\center
\begin{tabular}{cc}
\includegraphics[trim=10 0 10 10,clip,width=0.35\textwidth]{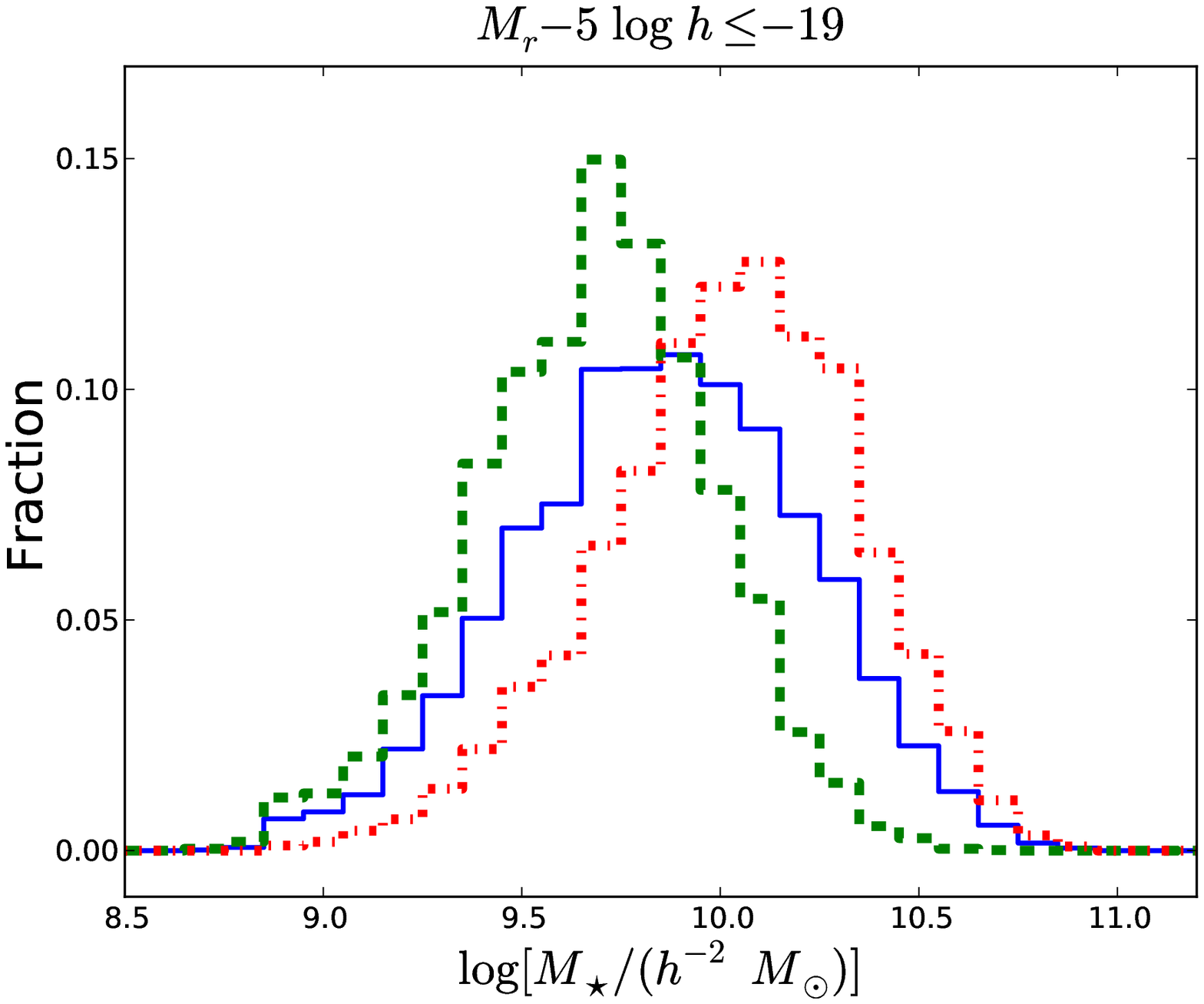}&
\includegraphics[trim=10 0 10 10,clip,width=0.35\textwidth]{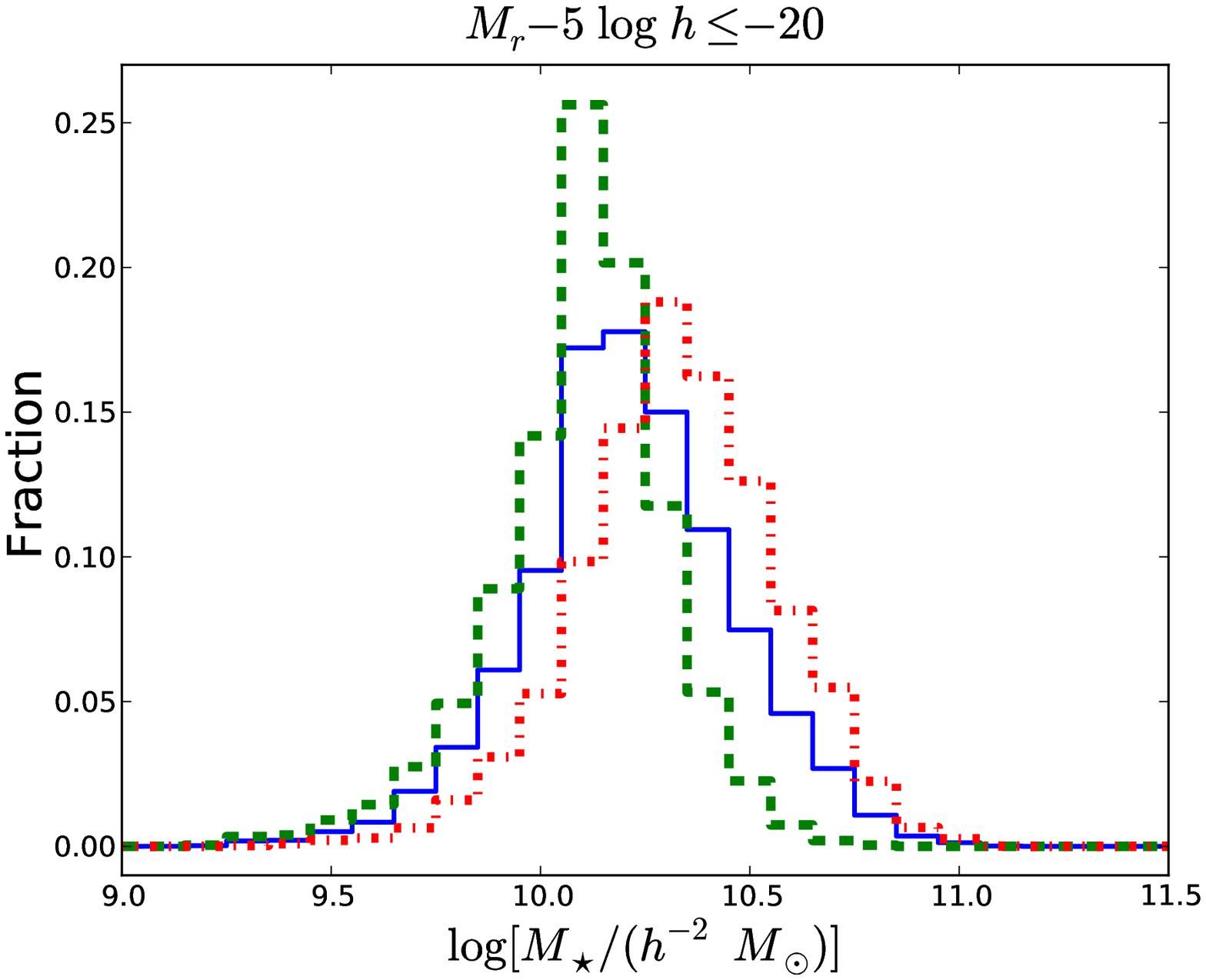}\\
\includegraphics[trim=10 0 10 37,clip,width=0.35\textwidth]{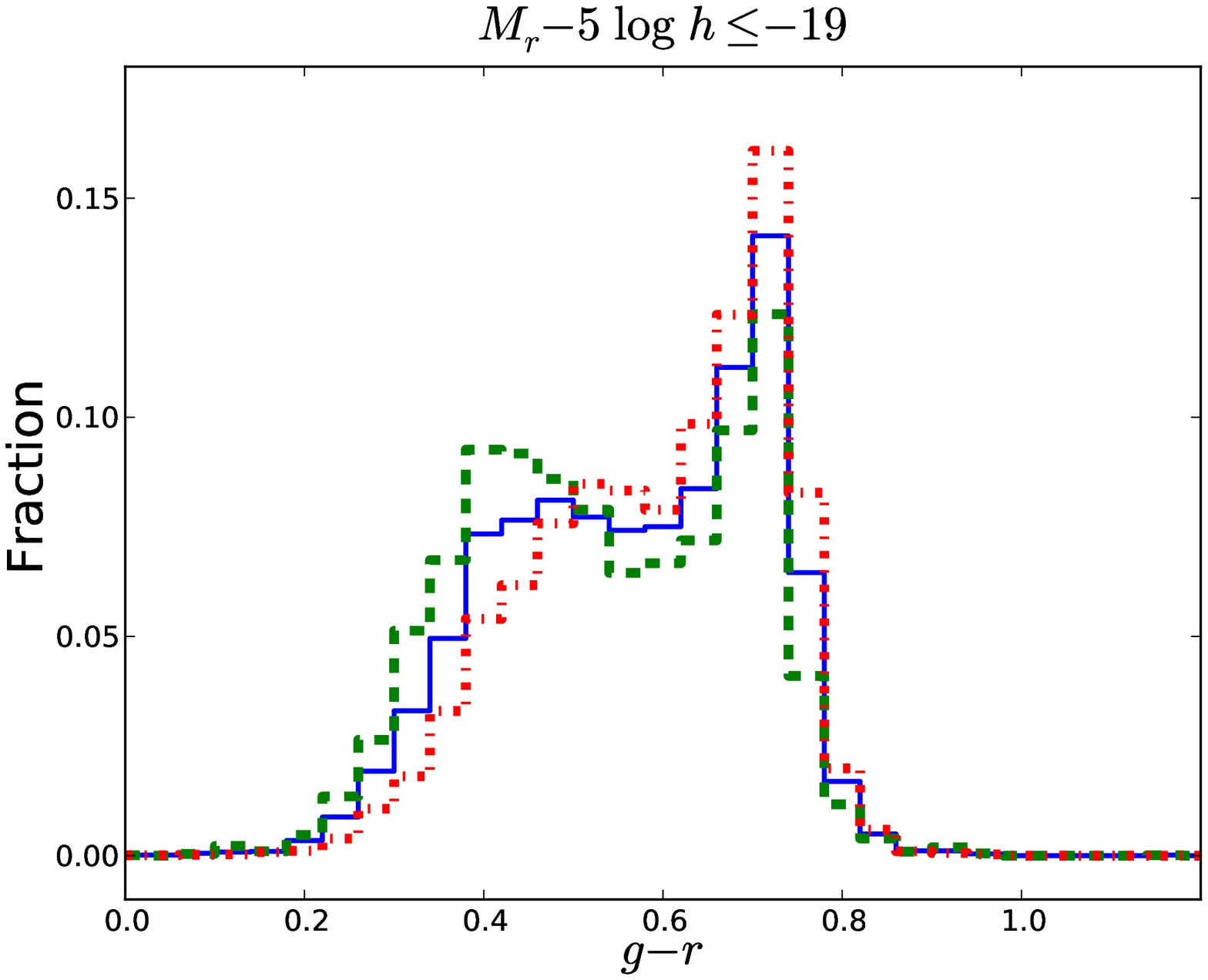} &
\includegraphics[trim=10 0 10 37,clip,width=0.35\textwidth]{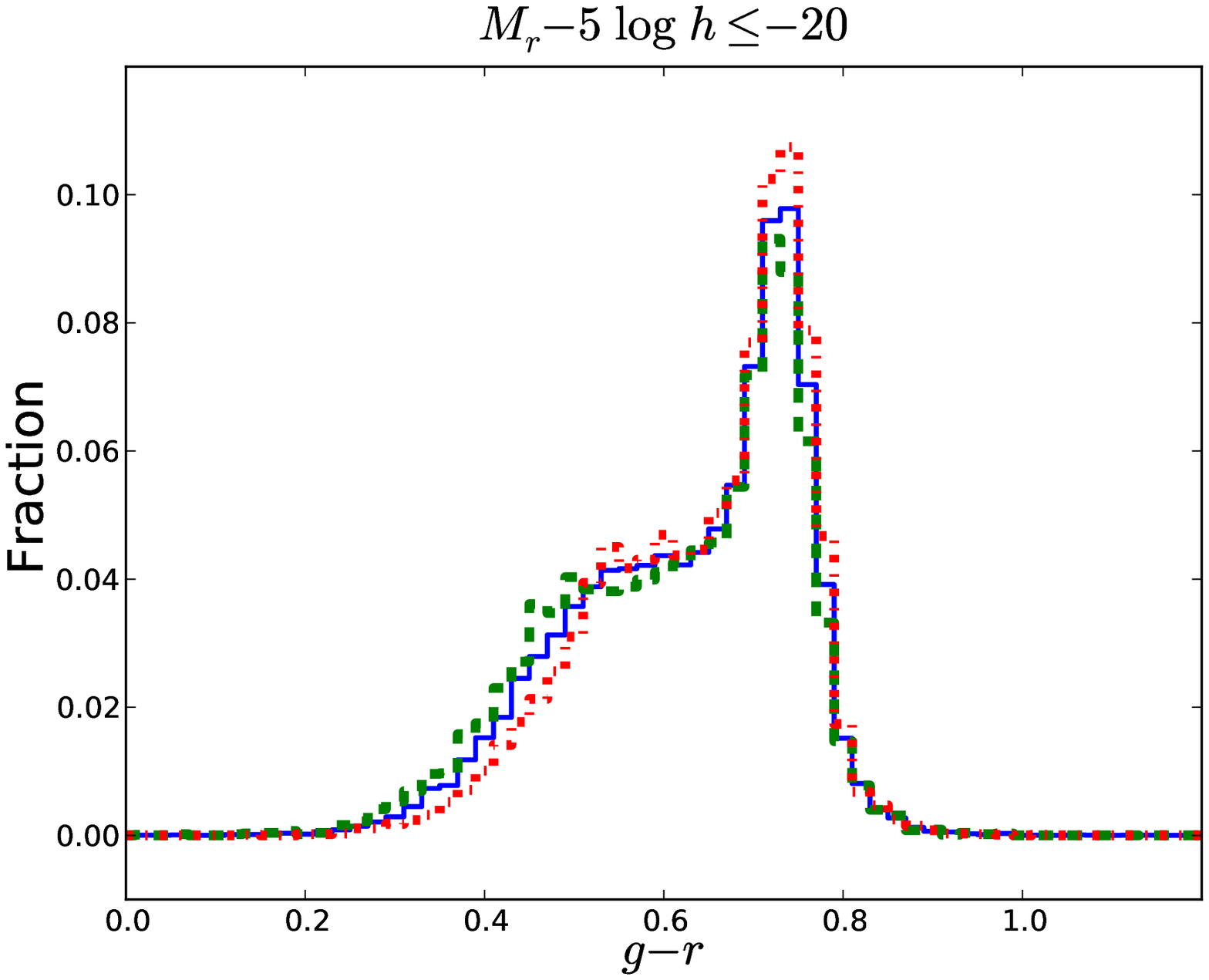}
\end{tabular}
\caption{\label{Fig:MstarNeq1SatCen} Stellar mass (top) and colour (bottom) distributions of the $-19$ (left) and $-20$ (right) $N=1$ populations contaminating the overall (solid blue), satellite (dashed green), and central (dot-dashed red) $N=2$ populations. Distributions are used with Equation (\ref{Eq:PureNeq2Hist}) to correct the $N=2$ populations for interloper systems.}
\end{figure*}

\begin{figure*}
\center
\begin{tabular}{cc}
\includegraphics[trim=5 36 40 30,clip,width=0.35\textwidth]{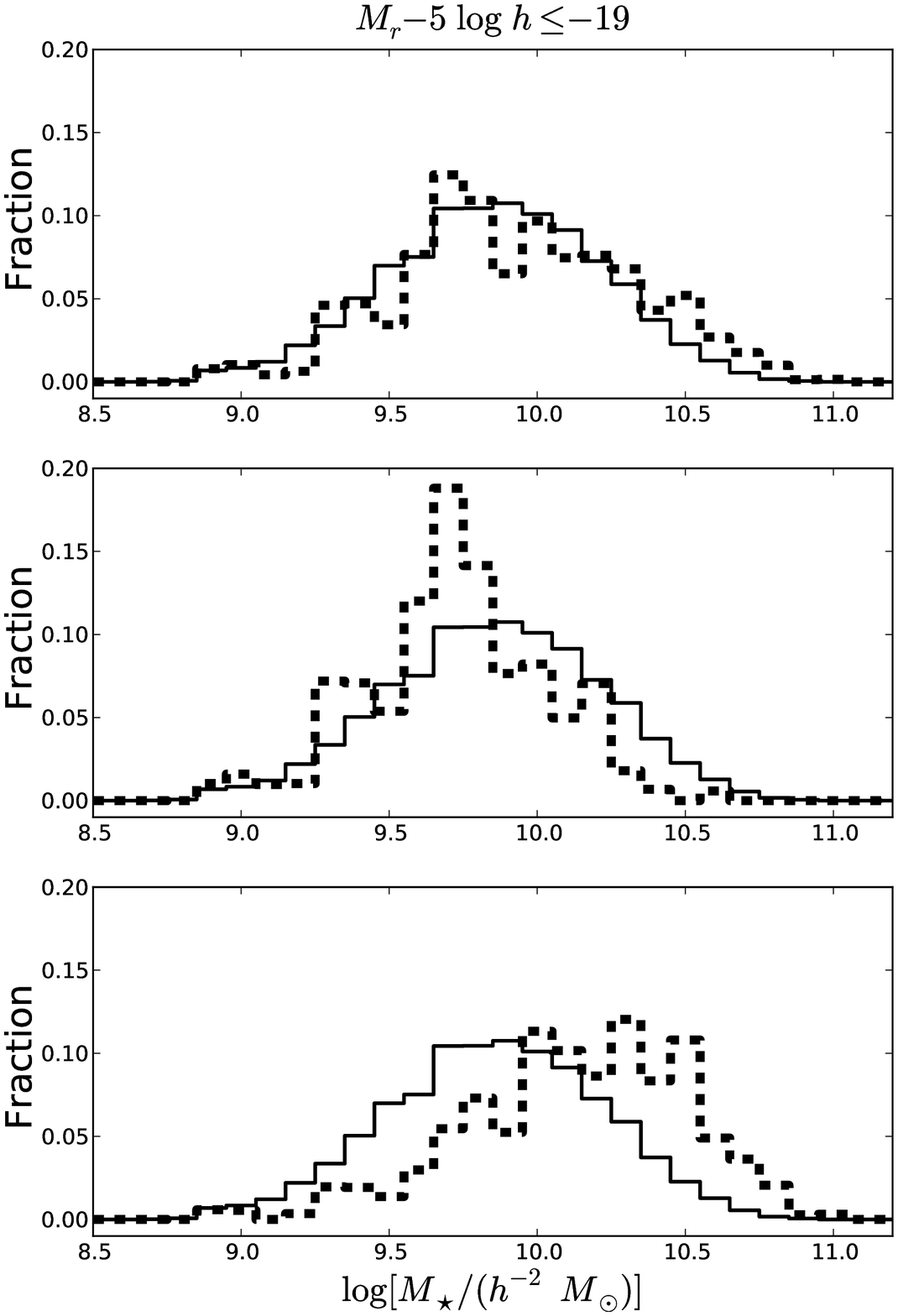} &
\includegraphics[trim=5 36 40 30,clip,width=0.35\textwidth]{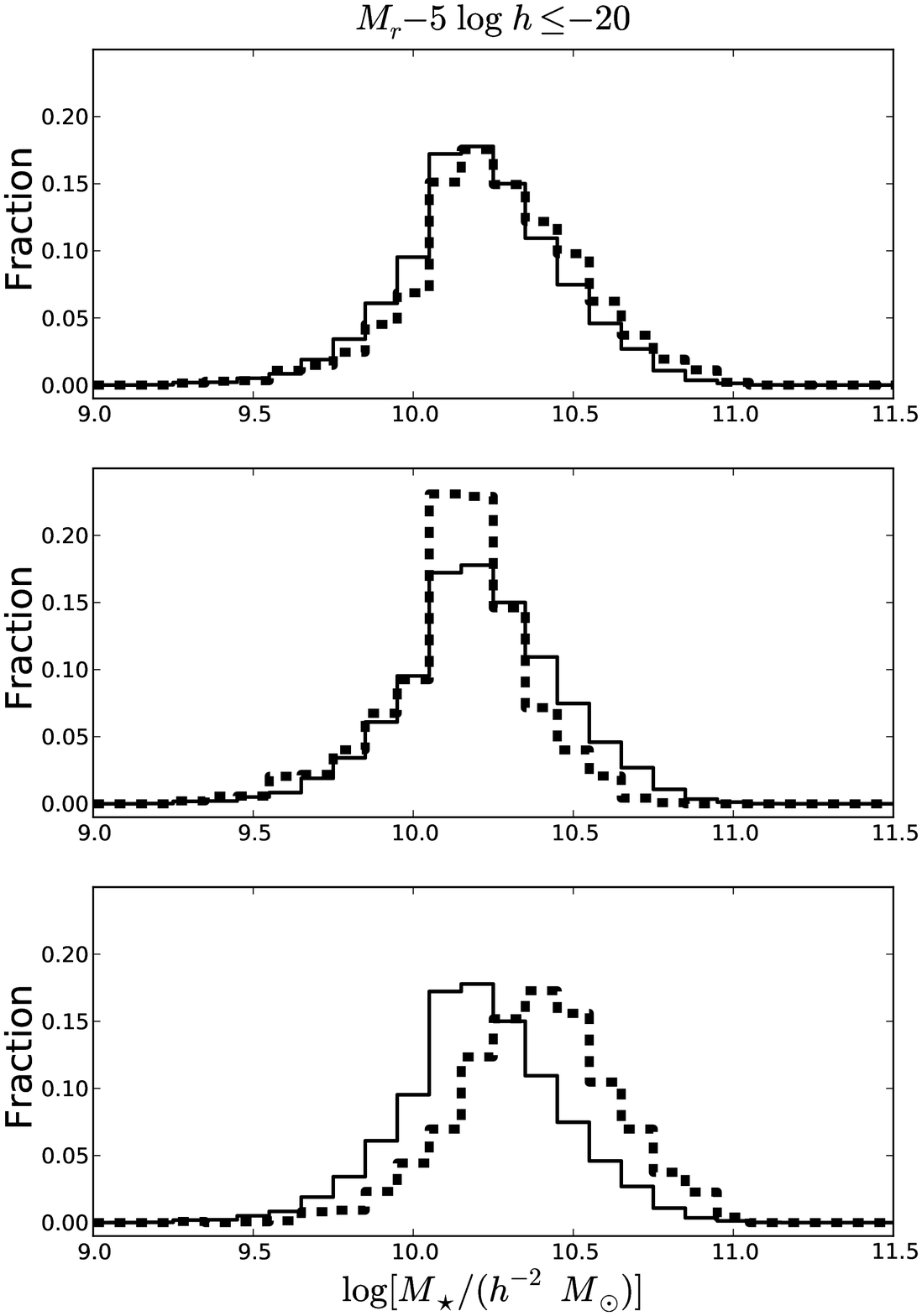}
\end{tabular}
\caption{\label{Fig:MstarNoResamp} Contamination-corrected stellar mass distributions of the $-19$ (left) and the $-20$ (right) overall (top), satellite (middle), and central (bottom) $N=2$ populations (dashed). Also shown is the stellar mass distribution of the full $N=1$ population (solid) without Monte Carlo mass resampling. Without Monte Carlo mass resampling, differences in stellar mass contribute to difference in colour between $N=1$ and $N=2$ populations.}
\end{figure*}

\subsection{Correction of $N=2$ populations for contamination by $N=1$ systems}\label{sec:contamination}
In the previous section, we observed that $N=2$ populations selected using our constraints on $D_{N}$ and $N_{700}$ are primarily contaminated by $N=1$ systems. Using our mock volume-limited redshift survey, we quantified the levels of contamination and the results are listed in Table \ref{sampletable}. Below we discuss our technique to statistically correct an $N=2$ population for contamination by $N=1$ systems. 

Let $H_{N=2}(x)$ be the observed distribution of galaxy property $x$ for an $N=2$ population selected using constraints on $D_{N}$ and $N_{700}$. The fraction of $N=2$ systems in this population is $f_{N=2}$ and the fraction of $N=1$ systems is $f_{N=1}$, as estimated from the mock volume-limited redshift survey. Thus,
\begin{equation}
H_{N=2}(x) = f_{N=1}*H_{N=1}^{\mathrm{pure}} (x)+ f_{N=2}*H_{N=2}^{\mathrm{pure}}(x),
\end{equation}
where $H_{N=i}^{\mathrm{pure}}(x)$ is the distribution for a pure $N=i$ population. We ignore the very small contamination by $N>2$ systems. $N=1$ populations selected by our constraints on $D_{N}$ and $N_{700}$ are $\ga99$ per cent pure and $H_{N=1}^{\mathrm{pure}}(x)$ is known. The distribution for a pure $N=2$ population may be approximated by
\begin{equation}\label{Eq:PureNeq2Hist}
H_{N=2}^{\mathrm{pure}}(x) \approx \frac{H_{N=2}(x) - f_{N=1}*H_{N=1}^{\mathrm{pure}}(x)}{f_{N=2} + f_{N>2}}.
\end{equation}

The satellite/central $N=2$ populations are also contaminated by a population of $N=1$ systems. The populations of $N=1$ systems contaminating the satellite/central $N=2$ populations are not the same as the population contaminating the overall $N=2$ population. To apply Equation (\ref{Eq:PureNeq2Hist}) to the satellite/central $N=2$ populations, the appropriate distributions, $H_{N=1}^{\mathrm{pure,cen}}(x)$ and $H_{N=1}^{\mathrm{pure,sat}}(x)$, must be determined. When a system identified using our constraints on $D_{N}$ and $N_{700}$ is two $N=1$ systems instead of an $N=2$ pair, the less\,(more) luminous $N=1$ galaxy is placed in the satellite\,(central) galaxy population. As a result, the $N=1$ population contaminating the satellite\,(central) $N=2$ population is less\,(more) luminous than the population contaminating the overall $N=2$ population. We construct the $N=1$ populations contaminating the satellite and central $N=2$ populations from the $N=1$ population contaminating the overall $N=2$ population by first randomly choosing two galaxies within the full $N=1$ population. The less\,(more) luminous galaxy is placed in the satellite\,(central) $N=2$ contaminant population. This process is repeated 10,000 times. Figure \ref{Fig:MstarNeq1SatCen} shows the stellar mass (top) and colour (bottom) distributions of the $-19$ (left) and $-20$ (right) $N=1$ populations contaminating the overall (solid blue), satellite (dashed green), and central (dot-dashed red) $N=2$ populations. 

\subsection{Stellar mass distribution and resampling}\label{sec:stellar}
Here, we examine the stellar mass distribution of our $N=2$ and $N=1$ galaxy populations and discuss our procedure for resampling these populations to be of the same stellar mass. The stellar mass distributions of the $-19$ (left) and $-20$ (right) overall (top), satellite (middle), and central (bottom) $N=2$ populations (dashed) are shown in Figure \ref{Fig:MstarNoResamp}. These distributions are contamination-corrected using Equation (\ref{Eq:PureNeq2Hist}) and the distributions shown in Figure \ref{Fig:MstarNeq1SatCen}. Also shown is the stellar mass distribution of the full $N=1$ population (solid).



From the stellar mass distributions, we see that satellite $N=2$ galaxies have less stellar mass when compared to $N=1$ galaxies, on average. On the other hand, central $N=2$ galaxies have more stellar mass when compared to $N=1$ galaxies, on average. Galaxies with higher stellar mass content tend to be redder in colour \citep{kauffmann2003}. Thus, differences in the colour distributions of our $N=2$ and $N=1$ populations are in part due to differences in stellar mass. 

We remove colour differences due to differences in stellar mass by comparing populations of the same stellar mass using a Monte Carlo technique to resample the much larger $N=1$ population, i.e. randomly selecting without replacement, subpopulations whose stellar mass distributions match the contamination-corrected $N=2$ distributions shown in Figure \ref{Fig:MstarNoResamp}. We check that the resampled $N=1$ population has the same stellar mass distribution as the $N=2$ population using the Kologorov-Smirnov (K-S) test probability computed using the function {\sc KolomogorovSmirnovTest} in {\sc Mathematica}. The K-S test probabilities when comparing the stellar mass of the $-19$ and $-20$ overall, satellite and central $N=2$ populations to the full and resampled $N=2$ populations are listed in Table \ref{table:kstable} for one realization of the Monte Carlo resampling. The resampled $N=1$ populations are clearly more like the various $N=2$ populations in stellar mass than the full $N=1$ population.



\begin{figure*}
\center
\begin{tabular}{cc}
\includegraphics[trim=10 30 50 30,clip,width=0.44\textwidth]{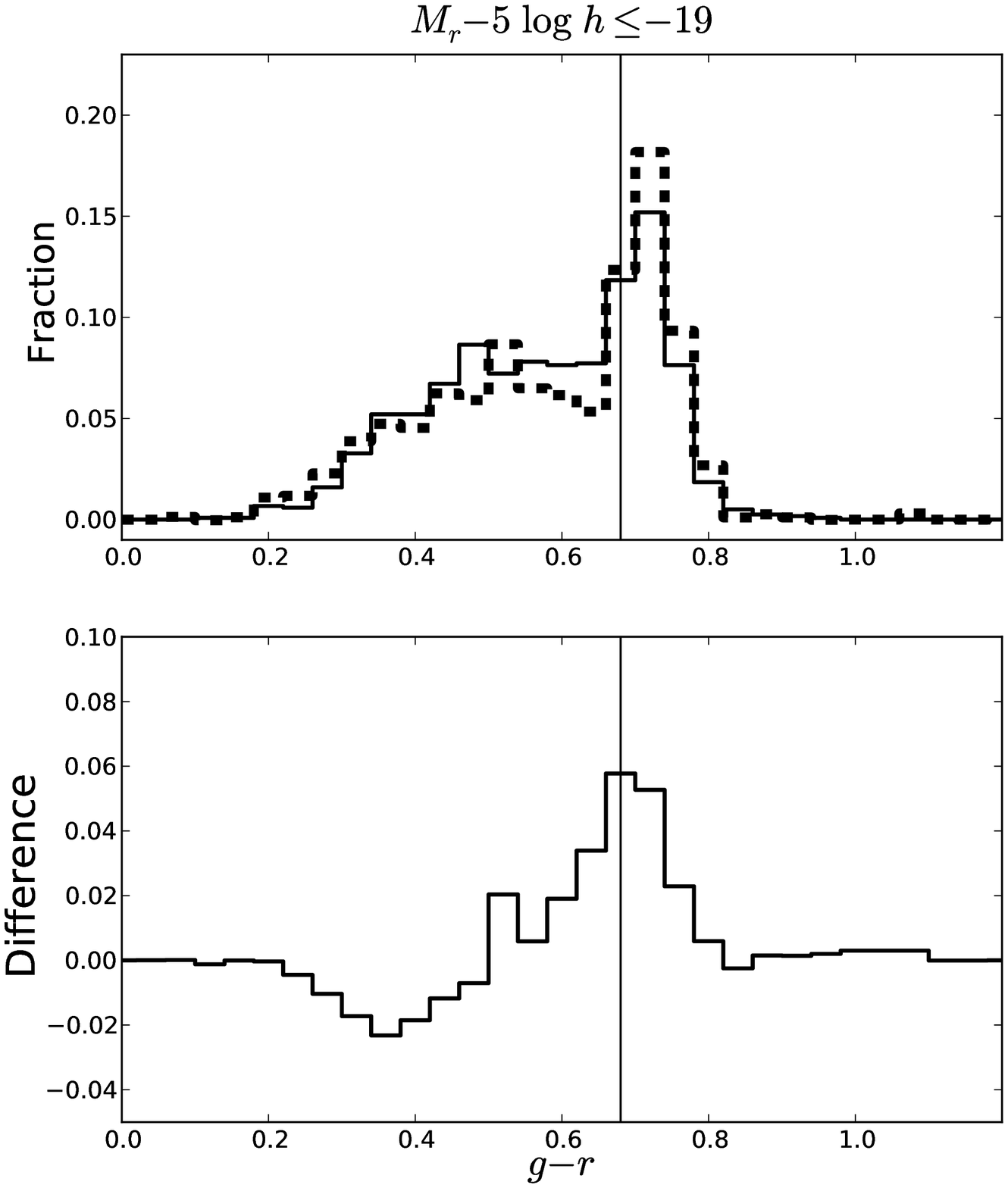} &
\includegraphics[trim=10 30 50 30,clip,width=0.44\textwidth]{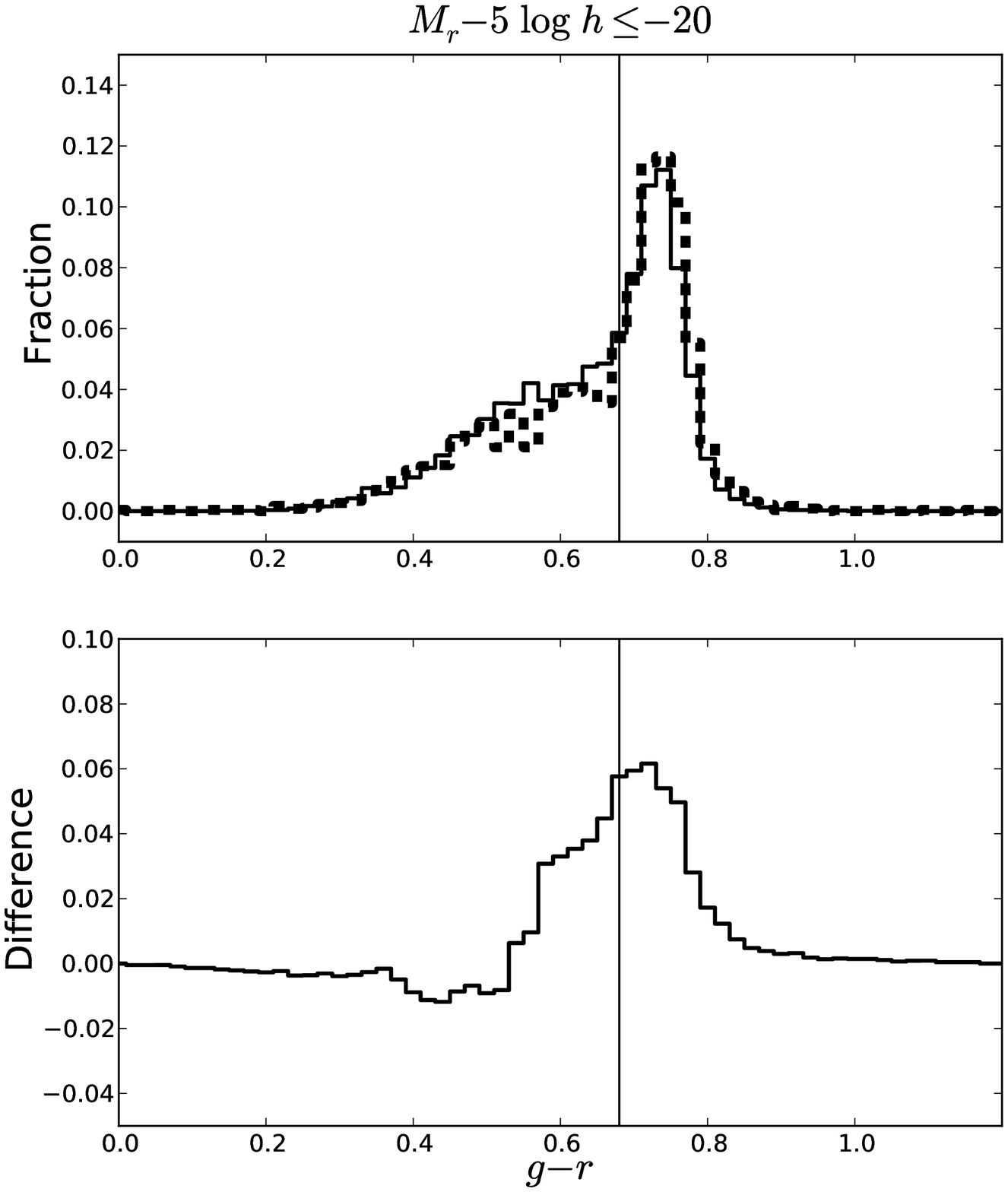}\\
\end{tabular}
\caption{\label{Fig:grPairs} (Top) Contamination-corrected $g-r$ distribution of the $-19$ (left) and $-20$ (right) overall $N=2$ population (dashed) and a resampled $N=1$ population (solid) of the same stellar mass. (Bottom) Difference between red fractions of the $-19$ (left) and $-20$ (right) $N=2$ and $N=1$ populations as a function of the $g-r$ value used to separate the red sequence and blue cloud. Black vertical line shows our chosen red/blue separator value at $g-r=0.68$ and the difference at this value is taken to be the red excess.}
\end{figure*}

\begin{figure*}
\center
\begin{tabular}{cc}
\includegraphics[trim=10 30 50 30,clip,width=0.44\textwidth]{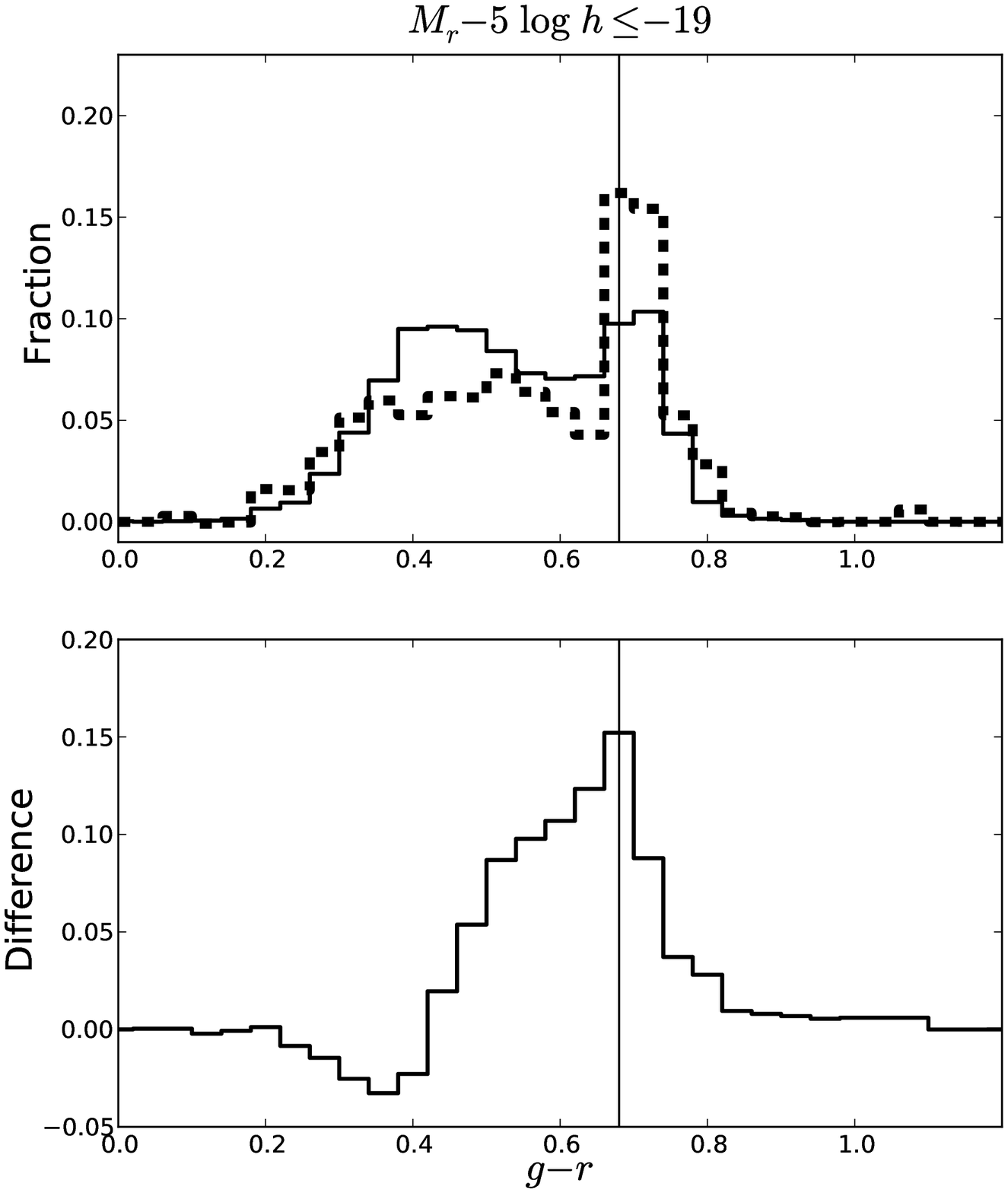} &
\includegraphics[trim=10 30 50 30,clip,width=0.44\textwidth]{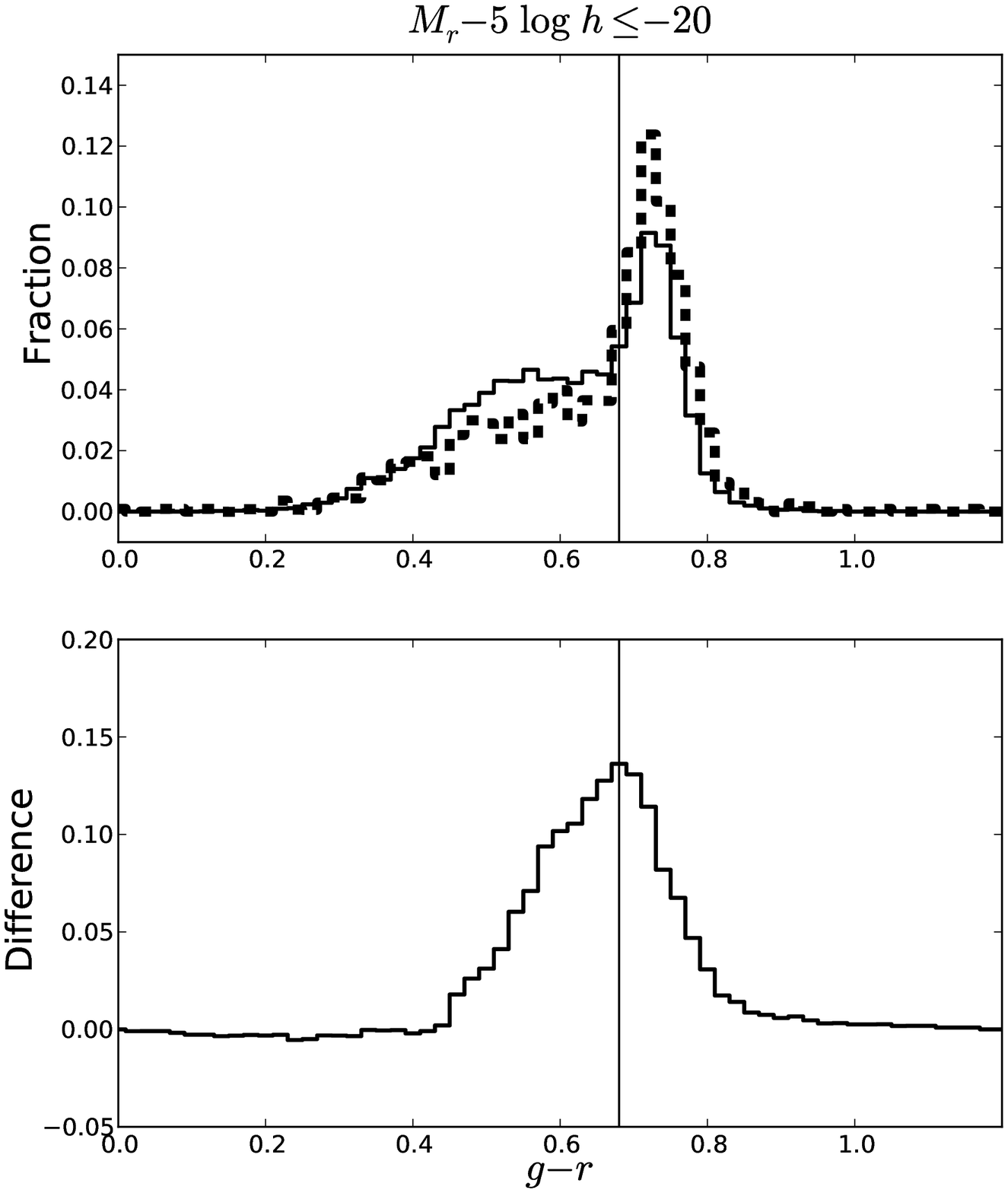}\\
\end{tabular}
\caption{\label{Fig:grSat} (Top) Contamination-corrected $g-r$ distribution of the $-19$ (left) and $-20$ (right) satellite $N=2$ population (dashed) and a resampled $N=1$ population (solid) of the same stellar mass. (Bottom) Difference between red fractions of the $-19$ (left) and $-20$ (right) satellite $N=2$ and $N=1$ populations as a function of the $g-r$ value used to separate the red sequence and blue cloud. Black vertical line shows our chosen red/blue separator value at $g-r=0.68$ and the difference at this value is taken to be the red excess.}
\end{figure*}

\begin{table}
 \centering
  \caption{Kolmogorov-Smirnov test probabilities when comparing the stellar mass of various $N=2$ populations with the full and resampled $N=1$ populations\label{table:kstable}}
  \begin{tabular}{@{}ll||llll@{}}
  \noalign{\smallskip}\hline
& Mag &  &  &  & \\
&Limit: &  \multicolumn{1}{l}{$-19$} & $-19$ &  \multicolumn{1}{l}{$-20$} & $-20$ \\
 \noalign{\smallskip}
& $N=1$: & Full & Resamp. & Full & Resamp. \\
 \noalign{\smallskip}\hline \noalign{\smallskip}
$N=2$ Overall& & 0.002 & 0.984 & 0.000 & 0.969 \\
$N=2$ Satellites & & 0.000 & 0.993 & 0.000 & 0.889 \\
$N=2$ Central & & 0.000 & 0.920 & 0.000 & 0.537 \\
 \noalign{\smallskip}\hline \noalign{\smallskip}
\end{tabular}
\end{table}

\section{Results}\label{sec:results}

\subsection{$N=2$ colour distribution}\label{sec:colour}
The goal of this paper is to investigate the differences in the colour between $N=1$ and $N=2$ galaxies and to discuss what these differences imply in regards to star formation activity. We compare the $g-r$ colour of $N=2$ populations with $N=1$ populations resampled to have the same stellar mass to remove any colour differences due to differences in stellar mass. NYU-VAGC magnitudes and colours are standard Petrosian magnitudes and colours \citep{petrosian1976,strauss2002} that are galactic extinction-corrected \citep{schlegel1998} and $K$-corrected by the template-fitting method of \citet{blanton2007}. 

The contamination-corrected $g-r$ distributions of the $-19$ (left) and $-20$ (right) overall $N=2$ populations (dashed) and the resampled $N=1$ populations (solid) of the same stellar mass are shown in the top row of Figure \ref{Fig:grPairs}. We separate galaxies into red (non-star-forming) and blue (star-forming) based on a cut in $g-r$ and measure the excess of red galaxies between the two populations. The difference between the red fractions of the $N=2$ and $N=1$ populations as a function of the $g-r$ value used to separate the red sequence and blue cloud is shown in the bottom row of Figure \ref{Fig:grPairs}.  A $g-r$ cut of 0.68 yields the maximum difference between the red fractions. This point corresponds to the ``green valley'' and we define a red galaxy to be a galaxy with $g-r \geq 0.68$ and a blue galaxy to be a galaxy with $g-r < 0.68$. This cut is shown as a vertical black line. The red excess is simply the difference between the two red fractions.

The $g-r$ distributions of the $N=1$ populations shown in Figure \ref{Fig:grPairs} change as the subpopulations selected during our Monte Carlo resampling changes. Thus, the red excess is different for each Monte Carlo realization. We generate 100 independent realizations of our Monte Carlo resampling technique and compute the red excess for each realization. Taking the mean and standard deviation over all realizations, we find that the overall $N=2$ population has a red excess of 0.05\,$\pm$\,0.01 and 0.06\,$\pm$\,0.01 for the $-19$ and $-20$ samples, respectively, relative to $N=1$ populations of the same stellar mass. Thus, $N=2$ galaxies in a very sparse group environment are redder than $N=1$ galaxies of the same stellar mass. This is not surprising considering that galaxies in denser group environments are well-known to be redder on average \citep[e.g.][]{weinmann2006a}. 


\subsection{Satellite $N=2$ red excess}\label{sec:satellites}
Following previous authors \citep[e.g.][]{vdb2008} we now study the satellite and central $N=2$ galaxy populations individually. The contamination-corrected $g-r$ distributions of the $-19$ (left) and $-20$ (right) satellite $N=2$ populations (dashed) and the resampled $N=1$ populations (solid) of the same stellar mass are shown in the top row of Figure \ref{Fig:grSat}. The difference in the red fractions of the satellite $N=2$ population and an $N=1$ population of same stellar mass as a function red/blue separator value is shown in the bottom row of Figure \ref{Fig:grSat}. As before we select $g-r=0.68$ as our red/blue separator value (shown by the black vertical line) and the red excess is the difference between the red fractions of the two populations. 

As before for the overall $N=2$ population, we generate 100 independent realizations of our Monte Carlo resampling technique and compute the red excess of the satellite $N=2$ population relative to $N=1$ populations of the same stellar mass for each realization. Taking the mean and standard deviation over all realizations, which accounts for the variation in the red fraction of the $N=1$ population, we find  that satellite $N=2$ populations have a red excess of 0.15\,$\pm$\,0.01 and 0.14\,$\pm$\,0.01 for the $-19$ and $-20$ samples, respectively, relative to $N=1$ populations of the same stellar mass. 

In other words, 15 per cent of satellite $N=2$ galaxies have transitioned from the blue to the red sequence after being accreted into a very sparse group of two halo, assuming $N=1$ galaxies are the progenitor of satellite $N=2$ galaxies in a statistical sense. Tracking the merger histories in $N$-body simulations reveals that the differences in past major merger ($<$3:1) histories of $N=1$ and $N=2$ subhaloes are $\la$1 per cent \citep{stewart2008}. Additionally, our cosmological model shows that for $D_{N} \leq$ 200\,$h^{-1}$\,kpc, the fraction of $N=2$ galaxies that have had a close pass within 30\,$h^{-1}$\,kpc within the last 0.5\,Gyr is $\sim$\,4 and 3 per cent for the $-19$ and $-20$ samples, respectively. Hence, mergers and galaxy-galaxy interactions are unlikely to play a significant role in these populations and the observed red excess is likely due to star formation quenching of the satellite galaxies from entering the sparse group environment as previously observed in richer and denser groups \citep[e.g.][]{weinmann2006a,vdb2008}.

Recent studies of satellite star formation quenching in galaxy groups reveal that quenching must occur over long time-scales of order 2--3\,Gyr \citep{kang2008,vdb2008,weinmann2009,mcgee2009,mcgee2011}. These authors suggest strangulation as the physical process responsible for star formation quenching in these systems. The hot halo gas of the infalling satellite galaxy is stripped (by ram-pressure or tides) and star formation slows as the cold gas is consumed. However, semi-analytic models show that instantaneous and complete removal of hot halo gas by ram-pressure stripping leads to a passive red fraction of satellite galaxies that is much higher than the observed fraction \citep{weinmann2006b,kang2008,kimm2009}. This has lead some authors to simply decrease the stripping efficiency in their semi-analytic model \citep[e.g.][]{font2008,weinmann2010} or suggest tidal stripping of hot halo gas \citep{weinmann2010} to better match the observations. 

However, strangulation of hot halo gas by any mechanism tends to produces too many galaxies in the green valley \citep{weinmann2010}. Because of this \citet{wetzel2012} do not support strangulation and point out that it is not clear that strangulation is efficient in low-mass haloes, such as the groups of two being studied here, as such haloes are not expected to have virial shock fronts which support hot, virialized gas within the halo \citep{dekel2006}. They also find a persistent specific star formation rate bimodality, i.e. the lack of galaxies in the green valley at all halo masses. Taken together,  they argue that the satellite quenching mechanism must bring about a rapid transition from the blue cloud to the red sequence, which any form of hot halo gas strangulation struggles with. Ram-pressure stripping of cold gas is the natural quenching mechanism for a rapid blue to red transition. Although most estimates indicate that ram-pressure stripping is very inefficient in the low-mass haloes being studied here, \citet{nichols2011} suggests that cold disc gas can be puffed up by internal star formation making the removal of cold gas by ram-pressure stripping possible even with a rarefied external medium of a low-mass halo. 

\begin{figure}
\center
\includegraphics[width=0.47\textwidth]{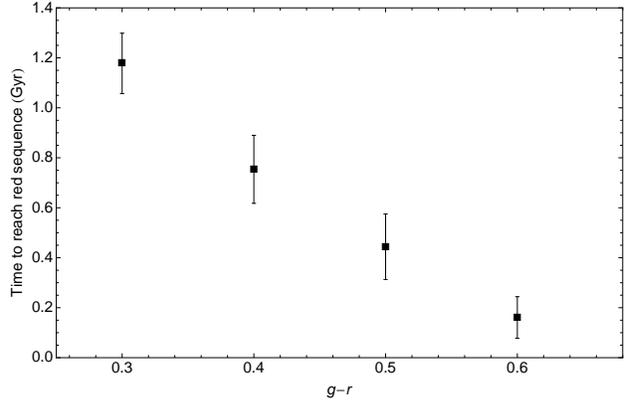}
\caption{\label{Fig:t2red} Time required to reach the red sequence after star formation stops (instantaneous and complete removal of cold gas) according to the population synthesis models of \citet{bc2003} using model galaxies with exponentially decaying star formation rates with $\tau=$1-10\,Gyr. Data points and error bars represent the mean and standard deviation over all values of $\tau$.\label{t2red}}
\end{figure}

Following the semi-analytic approach, we include a simple treatment of the removal of the cold gas from the infalling satellite galaxies (possibly by ram-pressure stripping) to our cosmological model. First, we consider an immediate-rapid quenching scenario where cold gas is instantaneously and completely removed immediately upon accretion. This is modelled using the population synthesis models of \citet{bc2003} with a \citet{chabrier2003} initial mass function with model galaxies with exponentially decaying star formation rates with $\tau = 1-10$\,Gyr. The infalling galaxy's star formation rate is set to zero when the galaxy reaches the ``initial'' $g-r$ value and we record the amount of time it takes to reach to the red sequence for each $\tau$. Figure \ref{Fig:t2red} shows the average time to reach the red sequence ($g-r \geq 0.68$) after star formation has stopped as a function of $g-r$ at accretion. The data points and error bars represent the mean and standard deviation over all values of $\tau$. 

Our analysis indicates that a galaxy with $g-r\sim 0.3$ will take an average of $\sim 1.2$\,Gyr after accretion to reach the red sequence in the immediate-rapid quenching scenario. Any galaxy with $g-r > 0.3$ will take less time. From the means and standard deviations in Figure \ref{Fig:t2red} we generate 100 realizations of the transition time, $t_{\mathrm{transition}}(g-r)$, for all $g-r$ bins between 0.3 and 0.68 by drawing from random distributions with the appropriate mean and standard deviation and then interpolating. 

\begin{figure}
\center
\includegraphics[trim=0 36 0 60,clip,width=0.5\textwidth]{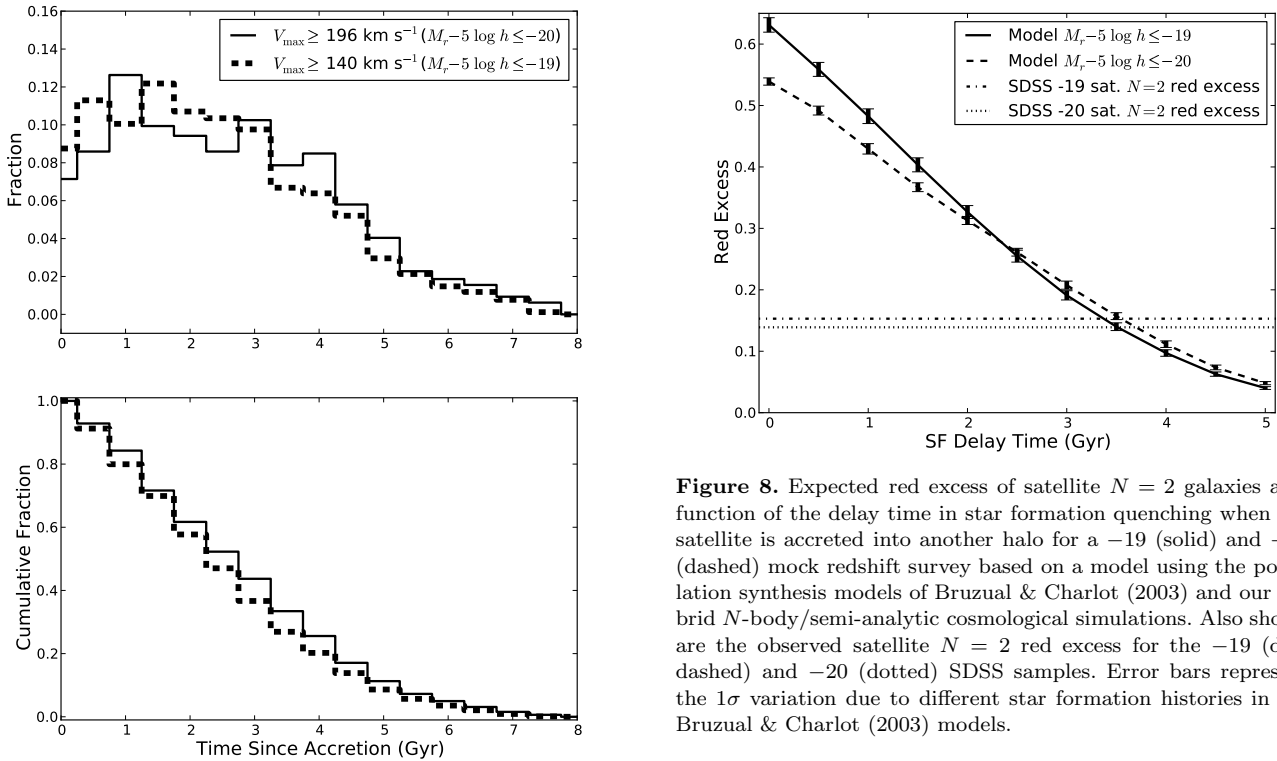}
\caption{\label{Fig:tacc} Time spent within host halo for simulated satellite $N=2$ galaxies with \Vmax $\geq$ 196\,km\,s$^{-1}$ (solid) and \Vmax $\geq$ 140 km s$^{-1}$ (dashed) in our hybrid $N$-body/semi-analytic cosmological model.}
\end{figure}

Next, we determine the transition fraction, $f_{\mathrm{transition}}(g-r)$, for all $g-r$ bins between 0.3 and 0.68. In the immediate-rapid quenching scenario this is simply the fraction that have been within their host halo for at least the transition time computed above. The distribution of time spent in host halo for satellite galaxies in our cosmological model, $f(t)$, is shown in the top panel of Figure \ref{Fig:tacc}. The bottom panel shows the right cumulative distribution, $F(t) = \sum_{t'\geq t}f(t')$, or the fraction that have been within their host halo for at least $t$\,Gyr. The transition fraction is given by $f_{\mathrm{transition}}(g-r) = F(t_{\mathrm{transition}}(g-r))$ and is computed for each $g-r$ bin between 0.3 and 0.68 for each realization. The satellite galaxies in our cosmological model have $N_{700}=1$, $D_{N}\leq$ 200\,$h^{-1}$\,kpc, and $N=2$. They are distinguished from central galaxies by a host/satellite flag tabulated in the catalog. We compute the look-back time at accretion for each satellite halo from the scale factor of the universe at accretion using $\Omega_{m}=0.3$, $\Omega_{\Lambda}=0.7$, $h=0.7$. 

To predict the red excess, we require the transition fraction estimated above and the $g-r$ distribution of a population of $N=1$ galaxies that become satellite $N=2$ galaxies. There is no reason to assume that $N=1$ galaxies of a particular colour are more likely to become a satellite $N=2$ galaxy. The $g-r$ distribution of this population should be identical to the $N=1$ population distribution shown in Figure \ref{Fig:grSat}, $H_{N=1}(g-r)$. The red excess is then given by
\begin{equation}
\textrm{red excess}=\left\langle \sum H_{N=1}(g-r) * f_{\mathrm{transition}}(g-r) \right\rangle,
\end{equation}
where the summation is over the range $0.3\leq g-r < 0.68$ and the average is over all 100 independent realizations. Our immediate-rapid quenching model predicts a red excess of 0.62\,$\pm$\,0.01 for the $-19$ sample and 0.54\,$\pm$\,0.01 for the $-20$ sample. The errors are taken to be the standard deviation over all 100 realizations. The red excess predictions of our immediate-rapid quenching model exceed the value observed in the NYU-VAGC DR6 samples by a factor of $\sim$\,4. 

\begin{figure}
\center
\includegraphics[trim=0 10 0 25,clip,width=0.5\textwidth]{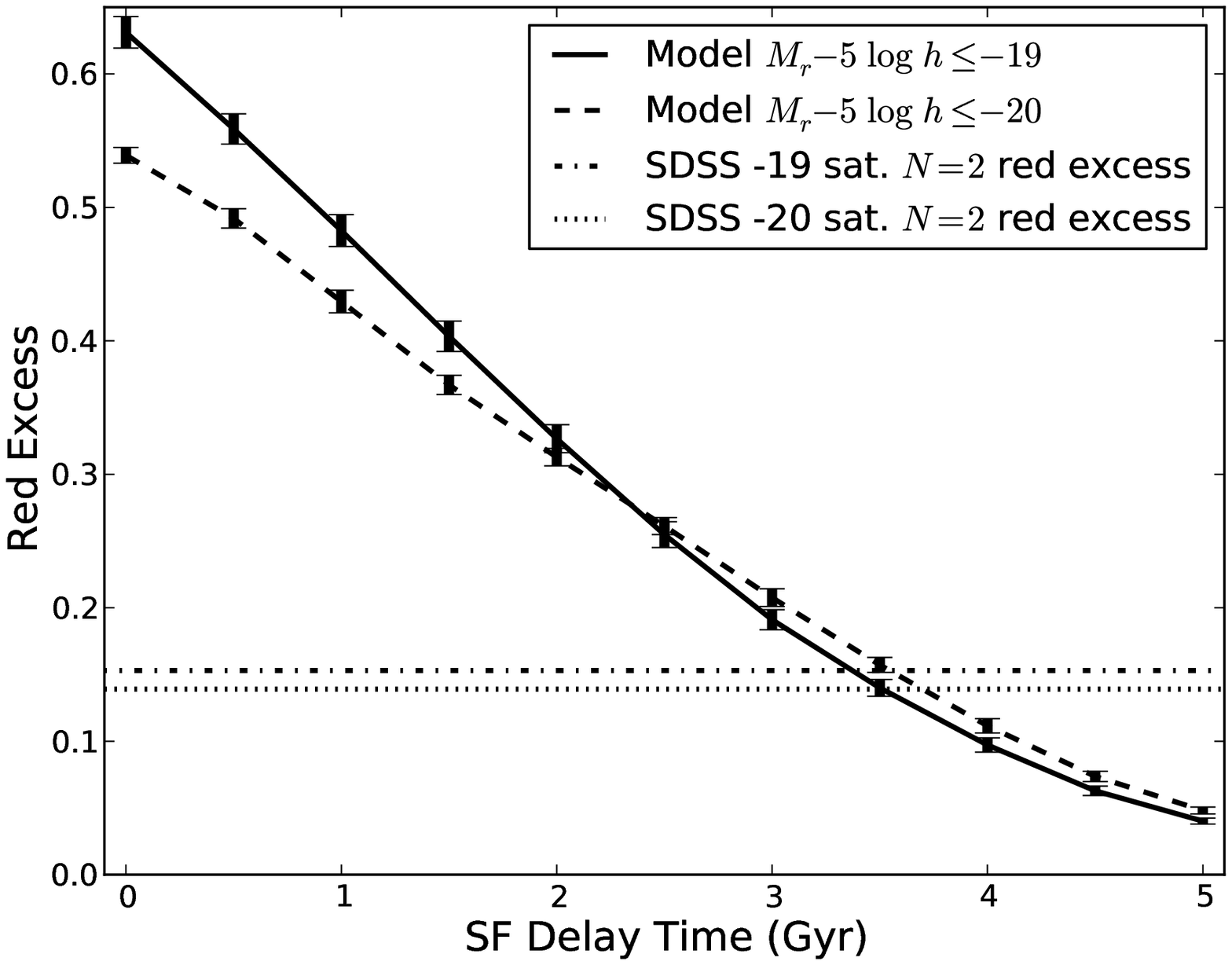}
\caption{\label{Fig:sf_delay} Expected red excess of satellite $N=2$ galaxies as a function of the delay time in star formation quenching when the satellite is accreted into another halo for a $-19$ (solid) and $-20$ (dashed) mock redshift survey based on a model using the population synthesis models of \citet{bc2003} and our hybrid $N$-body/semi-analytic cosmological simulations. Also shown are the observed satellite $N=2$ red excess for the $-19$ (dot-dashed) and $-20$ (dotted) SDSS samples. Error bars represent the 1$\sigma$ variation due to different star formation histories in the \citet{bc2003} models.}
\end{figure}

\begin{figure*}
\center
\begin{tabular}{cc}
\includegraphics[trim=10 30 50 30,clip,width=0.44\textwidth]{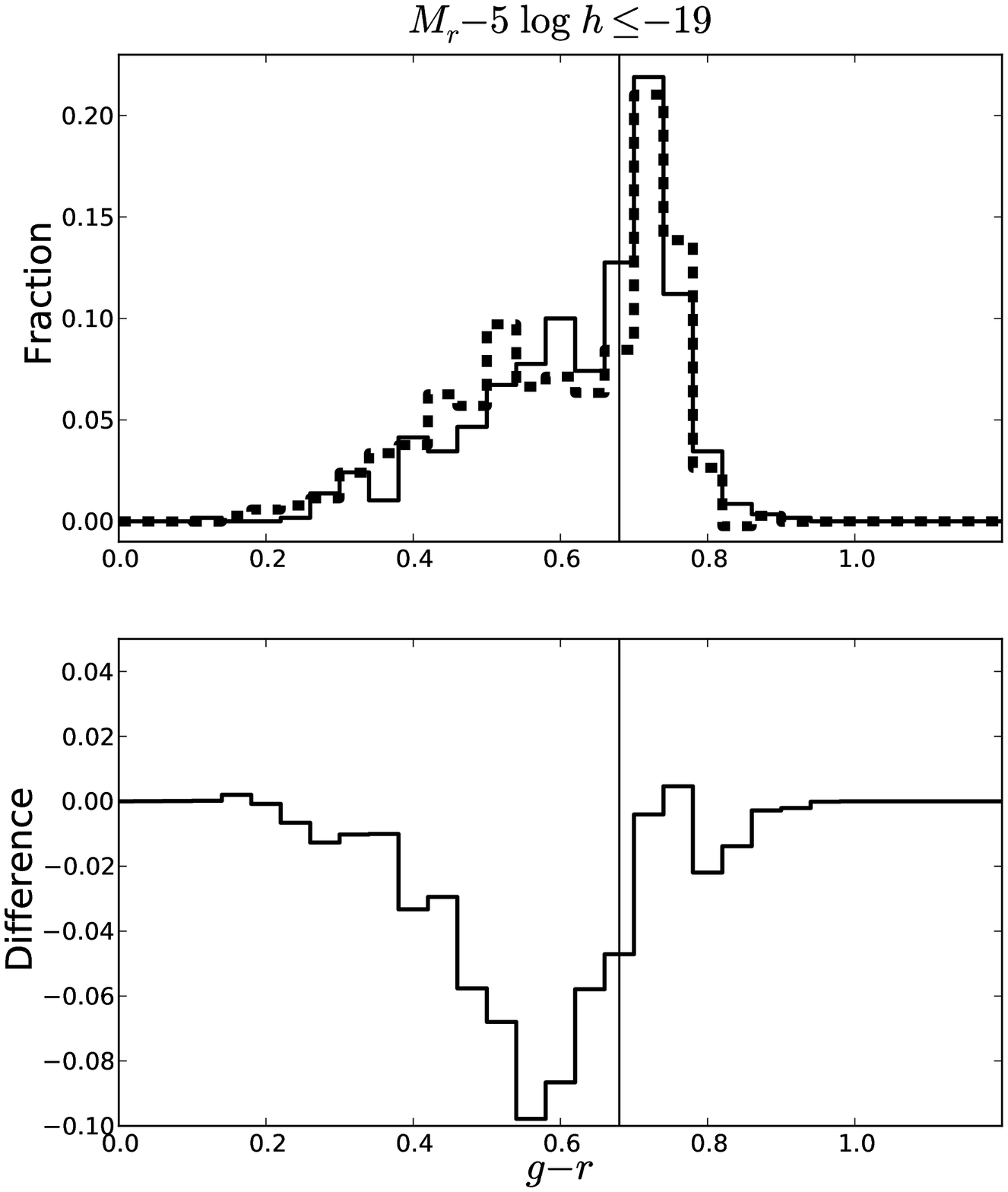} &
\includegraphics[trim=10 30 50 30,clip,width=0.44\textwidth]{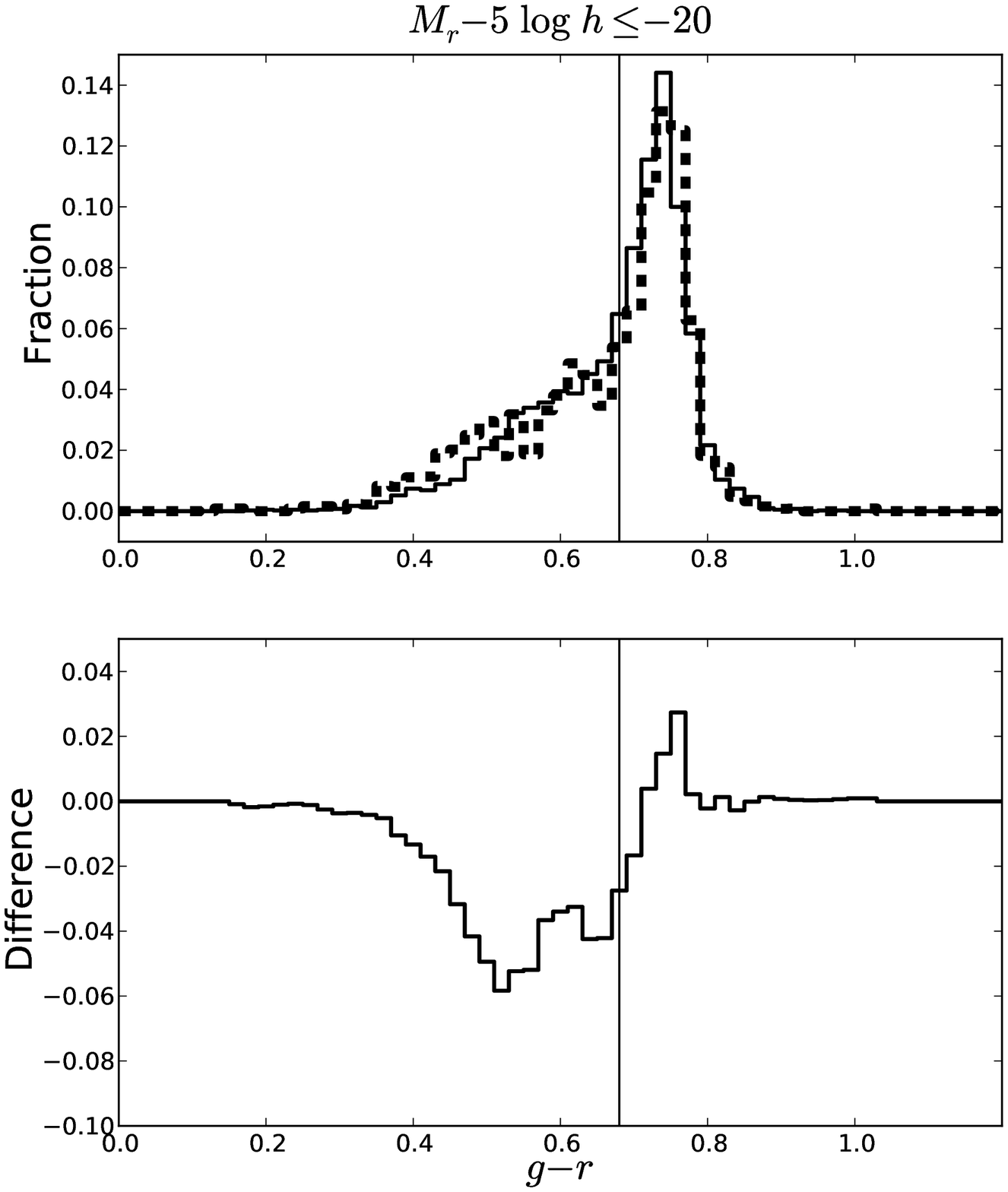}\\
\end{tabular}
\caption{\label{Fig:grCen} (Top) Contamination-corrected $g-r$ distribution of the $-19$ (left) and $-20$ (right) central $N=2$ population (dashed) and a resampled $N=1$ population (solid) of the same stellar mass. (Bottom) Difference between red fractions of the $-19$ (left) and $-20$ (right) central $N=2$ and $N=1$ populations as a function of the $g-r$ value used to separate the red sequence and blue cloud. Black vertical line shows our chosen red/blue separator value at $g-r=0.68$ and the difference at this value is taken to be the red excess.}
\end{figure*}

While \citet{wetzel2012} argue that the quenching mechanism must be rapid, it does not need to occur immediately upon accretion, i.e. a delayed-then-rapid scenario. We can easily model such a scenario by including a delay time before truncating star formation in our model galaxies. The time to reach the red sequence (see Figure \ref{Fig:t2red}) becomes $t_{\mathrm{transition}} \rightarrow t_{\mathrm{transition}} + t_{\mathrm{delay}}$. As the transition times get longer, the transition fractions decrease and the computed red excess decreases. The red excess predicted from the model as a function of star formation quenching delay time is shown in Figure \ref{Fig:sf_delay}. A delay time of $\sim$\,3.3 and 3.7\,Gyr predicts a red excess in agreement with the observations for the $-19$ and $-20$ samples, respectively.

A few caveats should be mentioned regarding the results above. Our observational result indicates that present-day satellite $N=2$ galaxies have a higher red fraction than present-day $N=1$ galaxies of the same stellar mass. In our immediate-rapid quenching model and delayed-then-rapid quenching model, we use a \emph{present-day} $N=1$ population to predict the red excess of \emph{present-day} satellite $N=2$ galaxies assuming the former are representative of the progenitors of the latter. Ideally, we would have used an $N=1$ population at the average redshift of accretion of satellite $N=2$ galaxies instead of the present-day population. From Figure \ref{Fig:tacc}, satellite $N=2$ galaxies have been within their host haloes for 2.3 and 2.6\,Gyr on average for the $-19$ and $-20$ samples, respectively. The average redshift of SDSS galaxies is $z\sim0.1$. Thus, we should be using an $N=1$ population at $z\sim$\,0.32 and 0.34 for the $-19$ and $-20$ samples, respectively \citep[see][]{mcgee2011}. \citet{vdb2008} also compared two present-day galaxy population while assuming one population is representative of the progenitors of the other. Following their arguments we do not expect significant evolution in the red fraction of $N=1$ galaxies at low redshifts and any evolution is probably towards lower red fractions at higher redshifts. As such, our model estimates are actually lower limits. 

The observed red excess of satellite $N=2$ galaxies with respect to $N=1$ galaxies of the same stellar mass may be biased by the $N=1$ population used to statistically correct the satellite $N=2$ population for contamination. If the red fraction of the pure $N=1$ population used in Equation (\ref{Eq:PureNeq2Hist}) is higher or lower, then the resulting red fraction of the satellite $N=2$ population and the resulting red excess will be lower or higher than the values reported above. According to our cosmological model, the average host halo mass of the satellite $N=2$ population is $\sim$\,0.4\,dex greater than the average host halo mass of the $N=1$ population selected using our constraints on $D_{N}$ and $N_{700}$ \citep[see Figure 4 of][]{barton2007}. This means that the satellite $N=2$ population is actually contaminated by an $N=1$ population with an average host halo mass greater than the $N=1$ population we used in Equation (\ref{Eq:PureNeq2Hist}) for statistical correction. Hence, we have slightly overestimated the true red fraction of the satellite $N=2$ population and therefore the red excess. Based on the dependence of the early-type fraction with host halo mass at fixed luminosity for central galaxies in groups as shown by \citet{weinmann2006a}, we estimate that the red fraction of the higher host halo mass $N=1$ population actually contaminating the satellite $N=2$ population is $\sim$\,0.1 greater than the red fraction of the $N=1$ population originally used in Equation (\ref{Eq:PureNeq2Hist}). If we increase the red fraction of the $N=1$ population by 0.1 and recompute the satellite $N=2$ red fraction using Equation (\ref{Eq:PureNeq2Hist}), we only find a decrease of $\sim$\,0.04 and 0.02 for $-19$ and $-20$, respectively. Thus, even when we account for the bias from the difference in host halo mass between the satellite $N-2$ population and the pure $N=1$ population used for statistical correction, we still find a red excess of at least $\sim$\,0.12. Thus, satellite star formation quenching is still present in $N=2$ systems and the slight bias from our statistical correction does not significantly change our conclusions regarding quenching time-scales (see Figure \ref{Fig:sf_delay}). 



The observed red excess may also be biased due to the effects of galaxy harassment by fainter satellites. $N=2$ haloes are expected to host more substructure suggesting that the effects of harassment by fainter satellites on satellite $N=2$ galaxies is greater than for $N=1 galaxies$, which implies that we have again slightly overestimated the red fraction of satellite N=2 galaxies. However, the impact of harassment in galaxy group appears to be secondary \citep{weinmann2006a} and we expect that accounting for harassment by fainter satellites will not have a significant impact on our observed red fraction.

Another assumption made in our simple model is that the orbits of infalling satellites are identical for satellites that are red or blue at accretion. If satellites that are red at accretion are biased towards particular orbits, then the distribution of time spent within the host halo may differ from that shown in Figure \ref{Fig:tacc} and our red excess estimates will be affected. It is difficult to tell whether the red excess estimates would be larger or smaller without ``a priori'' information on which orbits red satellites prefer. 

Finally, in this preliminary study we have treated gas removal as instantaneous and complete and we have not included AGN feedback in contrast to more sophisticated semi-analytic galaxy formation models such as \citet{kang2008}, \citet{font2008} and \citet{weinmann2010}. Nevertheless, our delayed-then-rapid model produces a star formation truncation time after accretion that is similar to the 3\,Gyr reported by \citet{mcgee2011} based on \citet{font2008}. In future work, we will studies of the properties of groups of two galaxies in considerably more depth and address the short-comings of our simple semi-analytic model in order to gain a more thorough understanding of the environmental process(es) at work in the very sparse galaxy groups.  

\begin{figure*}
\center
\begin{tabular}{cc}
\includegraphics[trim=10 30 50 30,clip,width=0.44\textwidth]{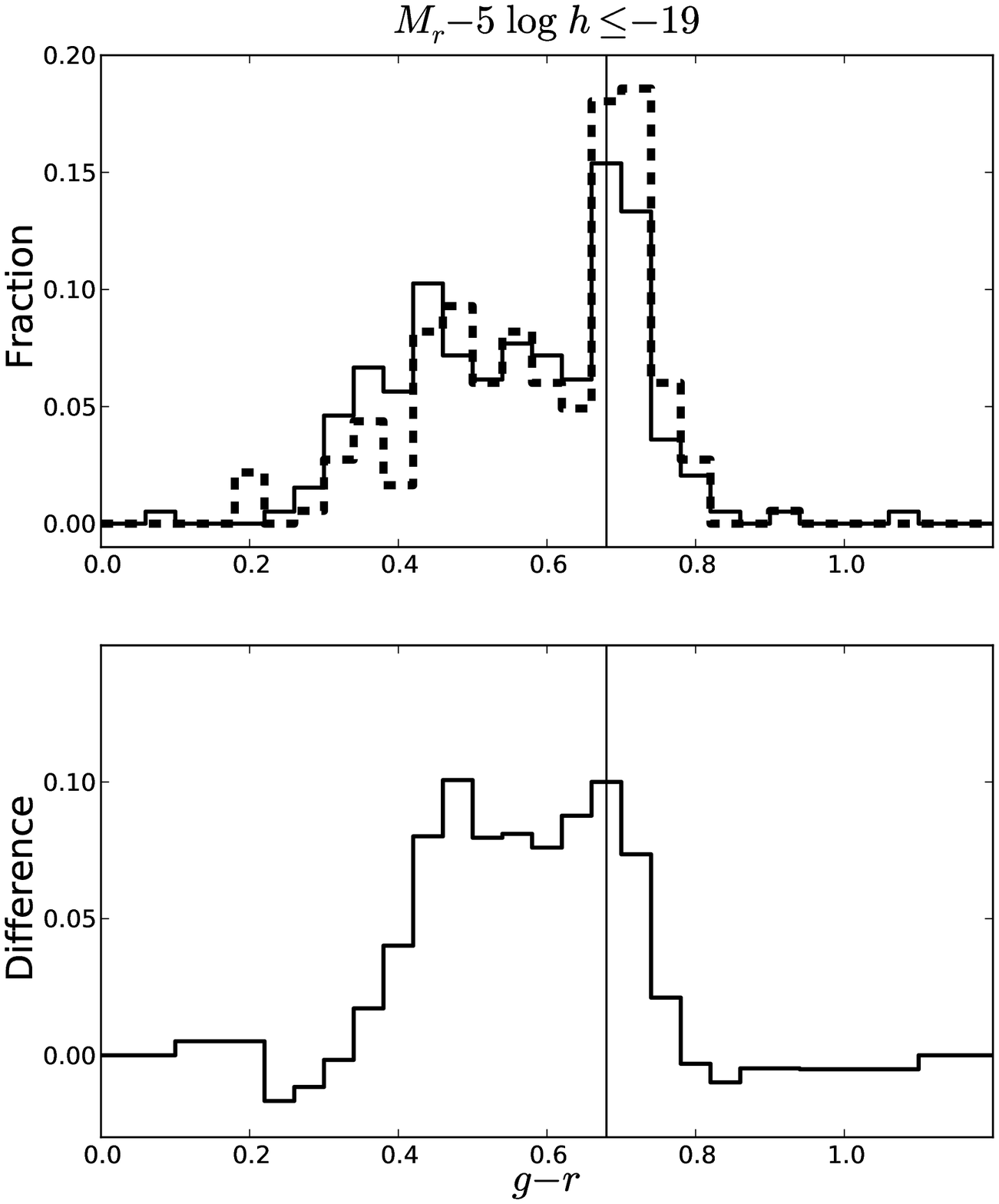} &
\includegraphics[trim=10 30 50 30,clip,width=0.44\textwidth]{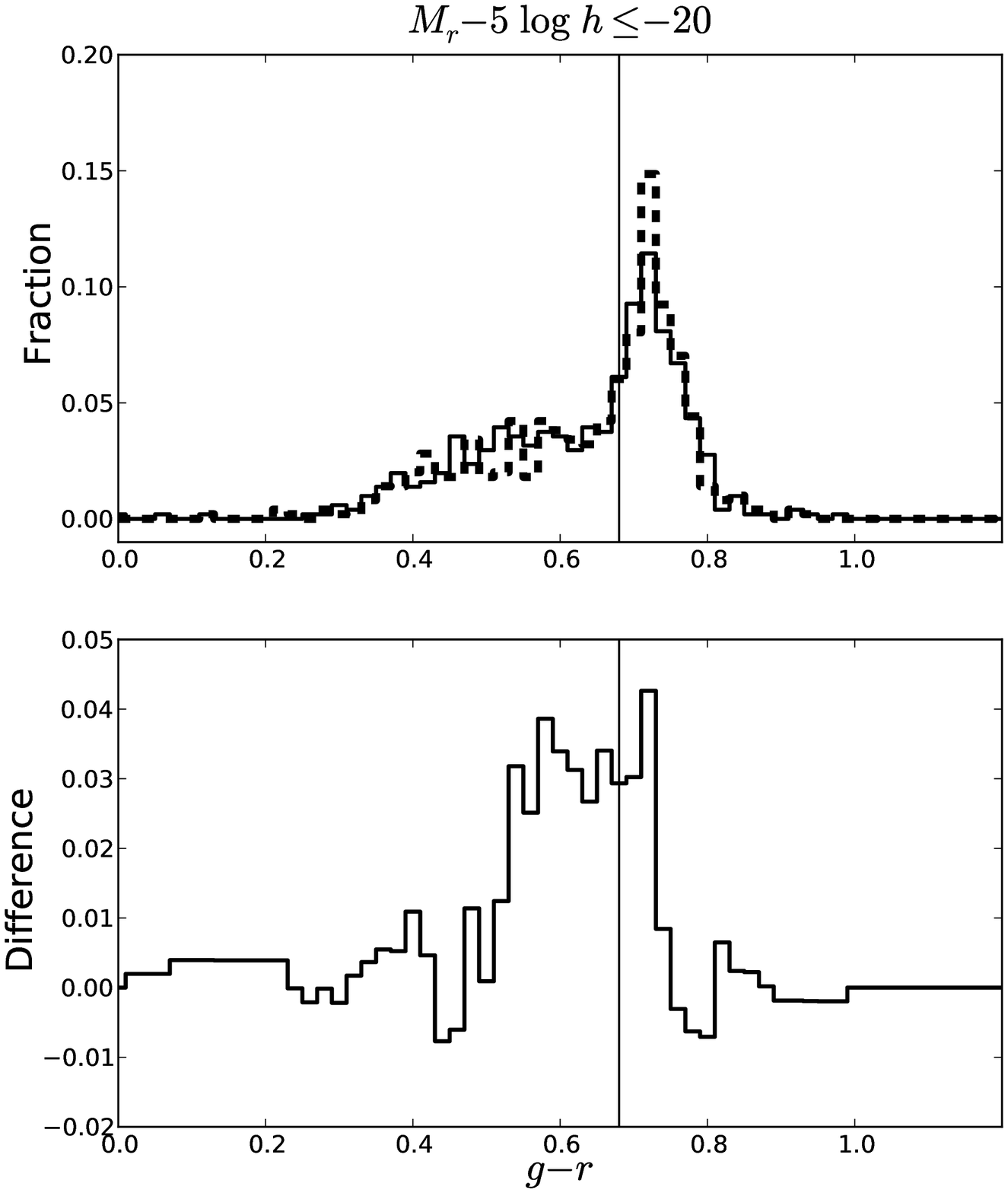}\\
\end{tabular}
\caption{\label{Fig:grSatRedCen} (Top) $g-r$ distribution of the $-19$ (left) and $-20$ (right) satellite $N=2$ population (dashed) with a red central galaxy and the overall satellite $N=2$ population (solid) of the same stellar mass from one realization of our Monte Carlo resampling technique. (Bottom) Difference between red fractions of the $-19$ (left) and $-20$ (right) satellite $N=2$ population with a red central and the overall satellite $N=2$ population of the same stellar mass as a function of the $g-r$ value used to separate the red sequence and blue cloud. Black vertical line shows our chosen red/blue separator value at $g-r=0.68$ and the difference at this value is taken to be the red excess.}
\end{figure*}

\begin{figure*}
\center
\begin{tabular}{cc}
\includegraphics[trim=10 30 50 30,clip,width=0.44\textwidth]{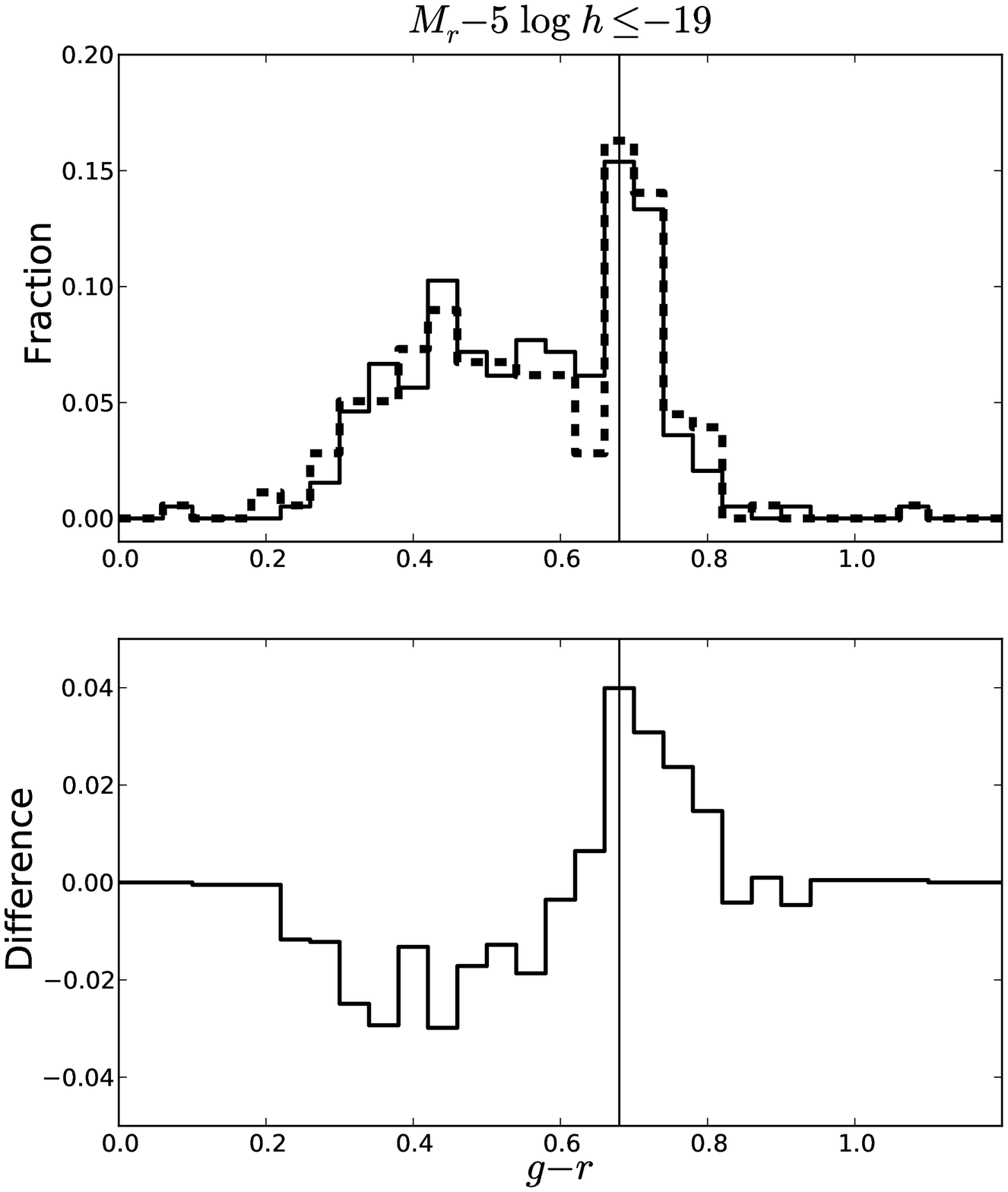} &
\includegraphics[trim=10 30 50 30,clip,width=0.44\textwidth]{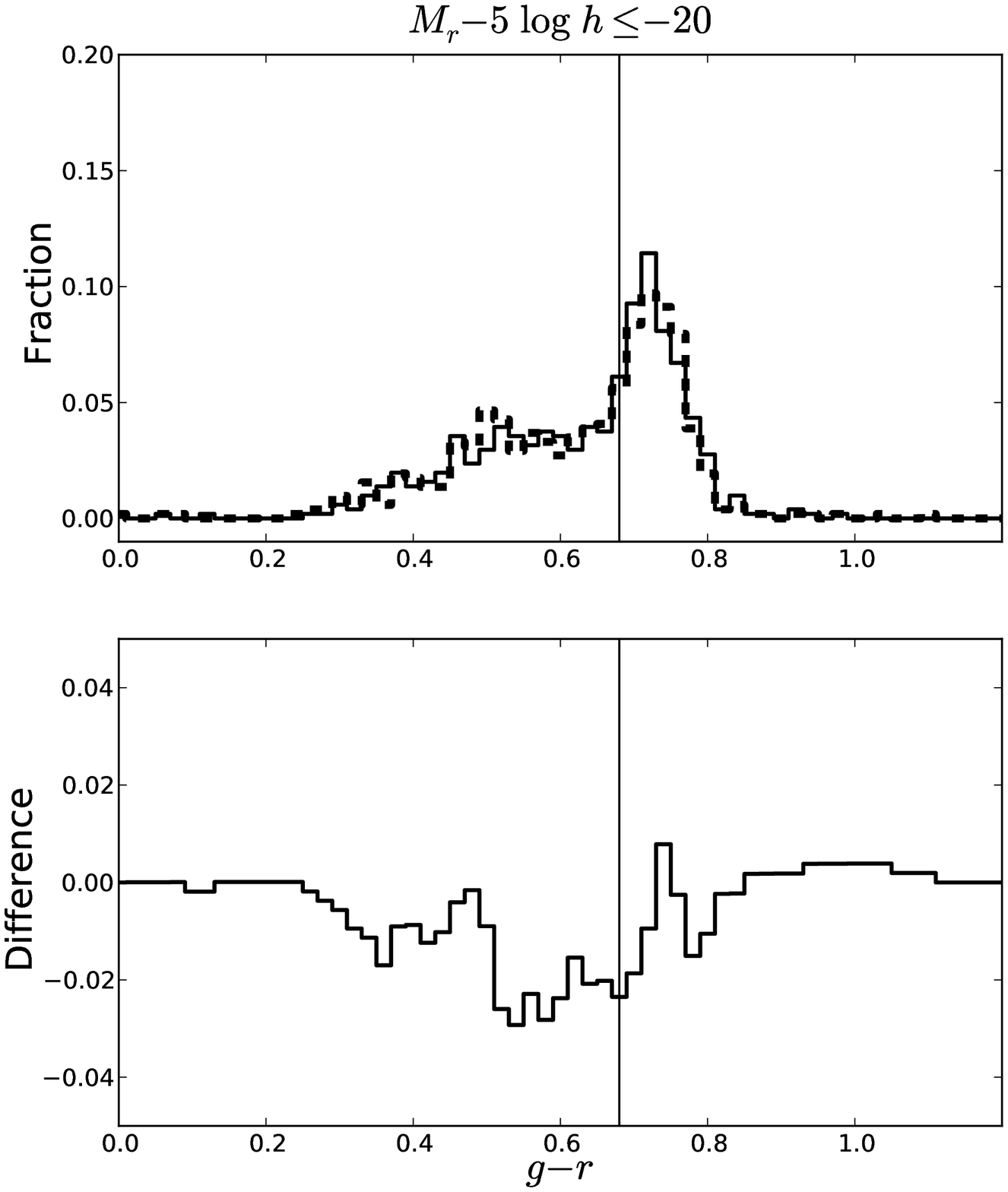} \\
\end{tabular}
\caption{\label{Fig:grSatBlueCen} (Top) $g-r$ distribution of the $-19$ (left) and $-20$ (right) satellite $N=2$ population (dashed) with a blue central galaxy and the overall satellite $N=2$ population (solid) of the same stellar mass from one realization of our Monte Carlo resampling technique. (Bottom) Difference between red fractions of the $-19$ (left) and $-20$ (right) satellite $N=2$ population with a blue central and the overall satellite $N=2$ population of the same stellar mass as a function of the $g-r$ value used to separate the red sequence and blue cloud. Black vertical line shows our chosen red/blue separator value at $g-r=0.68$ and the difference at this value is taken to be the red excess.}
\end{figure*}

\subsection{Central $N=2$ blue excess}\label{sec:centrals}
In the previous section we explored our satellite $N=2$ populations; we turn our attention to our central $N=2$ populations here. The contamination-corrected $g-r$ distributions of the $-19$ (left) and $-20$ (right) central $N=2$ populations (dashed) and the resampled $N=1$ populations (solid) of the same stellar mass are shown in the top row of Figure \ref{Fig:grCen}. The difference in the red fractions of the central $N=2$ population and an $N=1$ population of the same stellar mass as a function of the $g-r$ value used to separate the red sequence and blue cloud is shown in the bottom row of Figure \ref{Fig:grCen}. As before we select $g-r=0.68$ as our blue/red separator value (shown by the black vertical line) and the red excess is the difference in the red fractions of the two populations.  For central $N=2$ populations we find a blue excess of 0.06\,$\pm$\,0.02 and 0.02\,$\pm$\,0.01 for the $-19$ and $-20$ samples, respectively, relative to $N=1$ populations of the same stellar mass from mean and standard deviation of 100 independent realizations of our Monte Carlo resampling technique. Thus, for our chosen red/blue separator, central galaxies in a group of two are bluer than isolated galaxies of the same stellar mass. 

The red excess in satellite galaxies is likely due to a lack of star formation from gas loss. The gas that is lost by the satellite may be fed directly to the central, triggering star formation. This transfer could explain the observed blue excess in centrals. To investigate this hypothesis further, we examine satellite galaxies that are paired with red\,(blue) central galaxies and compare them with the overall satellite population. We begin by considering the stellar mass distributions of the three satellite populations: all satellites, satellites with a red central, and satellites with a blue central. These distributions are not contamination-corrected using Equation (\ref{Eq:PureNeq2Hist}) because we do not know the appropriate distributions for correction. The three satellite populations are resampled to be of the same stellar mass using our Monte Carlo resampling technique and we generate 100 independent realizations.

For each realization, we compute the $g-r$ distribution and red fraction for all three satellite populations. Figures \ref{Fig:grSatRedCen} and \ref{Fig:grSatBlueCen} show the results for satellites with a red and blue central, respectively, compared to the overall satellite population for one realization. The red excess of the satellite population with a red\,(blue) central galaxy relative to the overall satellite population is computed for each realization. From the mean and standard deviation over all realizations, satellites with a red central have a red excess of of 0.05\,$\pm$\,0.03 and 0.02\,$\pm$\,0.02 for the $-19$ and $-20$ samples, respectively, relative to the overall satellite population of the same stellar mass. Similarly, the population of satellites with a blue central have a blue excess of 0.00\,$\pm$\,0.03 and 0.03\,$\pm$\,0.02 for the $-19$ and $-20$ samples, respectively, relative to the overall satellite population of the same stellar mass. In other words, red satellites slightly tend to pair with red centrals and blue satellites slightly tend to pair with blue centrals and ``galactic conformity'' \citep{weinmann2006a} is somewhat present even in groups of two. However, we conclude from this analysis that direct gas exchange between the satellite and central is not a likely cause of the central blue excess.

Triggered star formation from a close satellite pass \citep[e.g.][]{mihos1996,barton2007} could also explain the observed blue excess in centrals, although we mentioned earlier that the frequency of such a close pass is expected to be very small. To test this hypothesis, we divide the central $N=2$ population into a sample with satellite separations $\la$\,100\,$h^{-1}$\,kpc and another with satellite separations $\ga$\,100\,$h^{-1}$\,kpc. This satellite separation value divides the central population into two nearly equal sized subsamples. The subsample with the closer satellites contains a red excess of 0.08\,$\pm$\,0.05 and blue excess of 0.03\,$\pm$\,0.03 for the $-19$ and $-20$ samples, respectively, relative to the subsample with wider satellite separations. Here, the errors are simply Poisson errors. We do not find that the central population with closer satellites to be bluer than the population with wider satellite separations. Hence, triggered star formation from a close pass is not a likely cause of the central blue excess either.

Lastly, we note that the central blue excess could also result because galaxies with satellites are also more likely to be actively accreting cold gas \citep[see][and references therein]{keres2009}, which is usually associated with larger-scale filamentary over-densities in galaxy position. However, we have no means of directly testing this scenario presently and we do not attempt to substantiate any further. 


\section{Conclusion}\label{sec:conclusion}
In this paper, we used a $\Lambda$CDM cosmological model of structure formation to devise a method for selecting galaxies that are isolated in their dark matter halo ($N=1$ system) and galaxies sharing their dark matter halo with exactly one neighbour ($N=2$ systems) in a group of exactly two based on $D_{N}$, the comoving projected distance to the nearest neighbour within $\Delta V \leq 1000$\,km\,s$^{-1}$ and $N_{700}$, the total number of galaxies within a comoving projected distance of 700\,$h^{-1}$\,kpc within $\Delta V\leq 1000$\,km\,s$^{-1}$. Our cosmological model enabled us to understand and correct for the contamination by galaxies in other environments allowing us to study the full, uncontaminated distributions of star-forming and morphological parameters instead of just the average trends. Using a Monte Carlo resampling technique, we constructed populations of isolated galaxies and groups of two galaxies with the same stellar mass distribution in order to remove any colour differences due to a difference in stellar mass. We studied the differences in $g-r$ to determine the effects of very sparse galactic group environments and we find the following:
\begin{enumerate}
\item{If galaxies are separated into a red sequence ($g-r\geq 0.68$) and blue cloud ($g-r < 0.68$), $N=2$ systems have a red excess of 0.05\,$\pm$\,0.01 and 0.06\,$\pm$\,0.01 for the $-19$ and $-20$ samples, respectively, relative to $N=1$ systems of the same stellar mass. Thus, the environment of even a sparse group environment influences galaxy evolution to a limited extent.}
\item{Examining the less luminous member, satellite $N=2$ galaxies have a red excess of 0.15\,$\pm$\,0.01 and 0.14\,$\pm$\,0.01 for the $-19$ and $-20$ samples, respectively, relative to $N=1$ galaxies of the same stellar mass.} 
\item{An immediate-rapid star formation quenching scenario, where the cold gas of the infalling satellite galaxy is instantaneously and completely removed immediately upon accretion into the sparse group of two halo yields a red excess prediction of 0.62\,$\pm$\,0.01 and 0.54\,$\pm$\,0.01 for the $-19$ and $-20$ samples, respectively. Thus, an immediate-rapid star formation quenching scenario is inconsistent with the observations.}
\item{A delayed-then-rapid star formation quenching scenario, as suggested by \citet{wetzel2012}, with a delay time of $\sim$\,3.3 and 3.7\,Gyr for the $-19$ and $-20$ samples, respectively, yields a red excess prediction in agreement with the observed red excess for satellite $N=2$ galaxies relative to $N=1$ galaxies of the same stellar mass.}
\item{Examining the more luminous member, central $N=2$ galaxies have a blue excess of 0.06\,$\pm$\,0.02 and 0.02\,$\pm$\,0.01 for the $-19$ and $-20$ samples, respectively, relative to $N=1$ galaxies of the same stellar mass.}
\item{Satellite $N=2$ galaxies with a red central have a red excess of 0.05\,$\pm$\,0.03 and 0.02\,$\pm$\,0.02 for the $-19$ and $-20$ samples, respectively, relative to the overall satellite population of the same stellar mass. Satellite $N=2$ galaxies with a blue central have a blue excess of 0.00\,$\pm$\,0.03 and 0.03\,$\pm$\,0.02 for the $-19$ and $-20$ samples, respectively, relative to the overall satellite population of the same stellar mass. Thus, red satellites slightly tend to pair with red centrals and blue satellites slightly tend to pair with blue centrals demonstrating that galactic conformity somewhat present even in groups of two galaxies. However, the central blue excess cannot be explained by a simple direct exchange of gas between the satellite and central.} 
\item{Central $N=2$ galaxies whose satellite separation is $\la$\,100\,$h^{-1}$\,kpc have a red excess of 0.08\,$\pm$\,0.05 and blue excess of 0.03\,$\pm$\,0.03 for the $-19$ and $-20$ samples, respectively, relative to central $N=2$ galaxies with satellite separations $\ga$\,100\,$h^{-1}$\,kpc. Thus, triggered star formation from a close pass is unlikely to cause the observed central blue excess. The central blue excess may be due to cold flows.} 
\end{enumerate}

Our most significant result in this preliminary study is that present-day satellite galaxies in a group of two have a higher red fraction compared to present-day isolated galaxies of the same stellar mass. Thus, star formation quenching of satellite galaxies by the yet undetermined dominant group environmental process occurs even in very sparse group environments where there are no other bright neighbours and galaxy-galaxy interactions with the central galaxy are unlikely. Taken together with our other results, we have demonstrated that environmental processes influence even the sparest groups of luminous galaxies. Further investigation of the star formation rate, AGN fraction, age, metallicity and concentrations of these simple systems and direct comparisons with predictions from $\Lambda$CDM and high-resolution hydrodynamical simulations will help undercover the causes of the trends seen here and contribute significantly towards a comprehensive understanding of galaxy evolution.


\section*{Acknowledgments}

The authors thank Joss Bland-Hawthorn, Alison Coil, Jeff Cooke, Ren\'ee Pelton, Billy Robbins, and Sanjib Sharma for helpful discussions and encouragement to complete the work. 

CQT gratefully acknowledges support by the National Science Foundation Graduate Research Fellowship under Grant No. DGE-1035963. EJB acknowledges support by NSF grant AST-1009999. CQT, EJB and JSB acknowledge support from the UC Irvine Center for Cosmology. 

Funding for the Sloan Digital Sky Survey (SDSS) has been provided by the Alfred P. Sloan Foundation, the Participating Institutions, the National Aeronautics and Space Administration, the National Science Foundation, the U. S. Department of Energy, the Japanese Monbukagakusho, and the Max Planck Society. 

The SDSS is managed by the Astrophysical Research Consortium (ARC) for the Participating Institutions. The Participating Institutions are The University of Chicago, Fermilab, the Institute for Advanced Study, the Japan Participation Group, The Johns Hopkins University, Los Alamos National Laboratory, the Max-Planck-Institute for Astronomy (MPIA), the Max-Planck-Institute for Astrophysics (MPA), New Mexico State University, University of Pittsburgh, Princeton University, the United States Naval Observatory, and the University of Washington.

\label{lastpage}


\begin{thebibliography}{}
\footnotesize

\bibitem[\protect\citeauthoryear{Adelman-McCarthy et al.}{2008}]{amc2008} Adelman-McCarthy J.~K. et al., 2008, ApJS, 175, 297 

\bibitem[\protect\citeauthoryear{Allam et al.}{2005}]{allam2005} Allam S.~S., Tucker D.~L., Lee B.~C., Smith J.~A., 2005, AJ, 129, 2062

\bibitem[\protect\citeauthoryear{Allgood et al.}{2006}]{allgood2006} Allgood B., Flores R.~A., Primack J.~R., Kravtsov A.~V., Wechsler R.~H., Faltenbacher A., Bullock J.~S., 2006, MNRAS, 367, 1781

\bibitem[\protect\citeauthoryear{Baldry et al.}{2006}]{baldry2006} Baldry I.~K., Balogh M.~L., Bower R.~G., Glazebrook K., Nichol R.~C., Bamford S.~P., Budavari T., 2006, MNRAS, 373, 469

\bibitem[\protect\citeauthoryear{Balogh et al.}{2000}]{balogh2000} Balogh M.~L., Navarro J.~F., Morris S.~L., 2000, ApJ, 540, 113 

\bibitem[\protect\citeauthoryear{Balogh et al.}{2002}]{balogh2002} Balogh, M.~L. et al., 2002, ApJ, 566, 123 

\bibitem[\protect\citeauthoryear{Balogh et al.}{2004}]{balogh2004a} Balogh M.~L. et al., 2004, MNRAS, 348, 1355 


\bibitem[\protect\citeauthoryear{Barton et al.}{2007}]{barton2007} Barton E.~J., Arnold J.~A., Zentner A.~R., Bullock J.~S., Wechsler R.~H., 2007, ApJ, 671, 1538 

\bibitem[\protect\citeauthoryear{Bell et al.}{2004}]{bell2004} Bell E.~F. et al., 2004, ApJ, 608, 752 

\bibitem[\protect\citeauthoryear{Berrier et al.}{2011}]{berrier2011} Berrier H.~D., Barton E.~J., Berrier J.~C., Bullock J. S., Zentner A. R., Wechsler R. H., 2011, ApJ, 726, 1 

\bibitem[\protect\citeauthoryear{Berrier et al.}{2006}]{berrier2006} Berrier J.~C., Bullock J.~S., Barton E.~J., Guenther H.~D., Zentner A.~R., Wechsler R.~H., 2006, ApJ, 652, 56 

\bibitem[\protect\citeauthoryear{Blanton \& Roweis}{2007}]{blanton2007} Blanton M.~R., Roweis S., 2007, AJ, 133, 734

\bibitem[\protect\citeauthoryear{Blanton et al.}{2003}]{blanton2003} Blanton M.~R. et al., 2003, ApJ, 594, 186

\bibitem[\protect\citeauthoryear{Blanton et al.}{2005a}]{blanton2005} Blanton M.~R. et al., 2005, AJ, 129, 2562 

\bibitem[\protect\citeauthoryear{Blanton et al.}{2005b}]{blanton2005b} Blanton M.~R., Eisenstein D., Hogg D.~W., Schlegel D.~J., Brinkmann J., 2005, ApJ, 629, 143 


\bibitem[\protect\citeauthoryear{Blumenthal et al.}{1984}]{blumenthal1984} Blumenthal G.~R., Faber S.~M., Primack J.~R., Rees M.~J., 1984, Nat, 311, 517 

\bibitem[\protect\citeauthoryear{Bruzual \& Charlot}{2003}]{bc2003} Bruzual G., Charlot S., 2003, MNRAS, 344, 1000

\bibitem[\protect\citeauthoryear{Bower et al.}{2006}]{bower2006} Bower R.~G., Benson A.~J., Malbon R., Helly J. C., Frenk C. S., Baugh C. M., Cole S., Lacey C. G., 2006, MNRAS, 370, 645 

\bibitem[\protect\citeauthoryear{Boylan-Kolchin et al.}{2009}]{boylankolchin2009} Boylan-Kolchin M., Springel V., White S.~D.~M., Jenkins A., Lemson G., 2009, MNRAS, 398, 1150


\bibitem[\protect\citeauthoryear{Chabrier}{2003}]{chabrier2003} Chabrier G., 2003, PASP, 115, 763

\bibitem[\protect\citeauthoryear{Chandrasekhar}{1943}]{chandrasekhar1943} Chandrasekhar S., 1943, ApJ, 97, 255 



\bibitem[\protect\citeauthoryear{Cole et al.}{2000}]{cole2000} Cole S., Lacey C.~G., Baugh C.~M., Frenk C.~S., 2000, MNRAS, 319, 168 



\bibitem[\protect\citeauthoryear{Cooper et al.}{2006}]{cooper2006} Cooper M.~C. et al., 2006, MNRAS, 370, 198 





\bibitem[\protect\citeauthoryear{Conroy et al.}{2006}]{conroy2006} Conroy C., Wechsler R.~H., Kravtsov A.~V., 2006, ApJ, 647, 201 

\bibitem[\protect\citeauthoryear{Cox et al.}{2006}]{cox2006} Cox T.~J., Jonsson P., Primack J.~R., Somerville R.~S., 2006, MNRAS, 373, 1013 

\bibitem[\protect\citeauthoryear{Croton et al.}{2006}]{croton2006} Croton D.~J. et al., 2006, MNRAS, 365, 11 


\bibitem[\protect\citeauthoryear{Dekel \& Birnboim}{2006}]{dekel2006} Dekel A., Birnboim Y., 2006, MNRAS, 368, 2


\bibitem[\protect\citeauthoryear{Di Matteo et al.}{2005}]{dimatteo2005} Di Matteo T., Springel V., Hernquist L., 2005, Nat, 433, 604 


\bibitem[\protect\citeauthoryear{Dressler et al.}{1997}]{dressler1997} Dressler A. et al., 1997, ApJ, 490, 577 

\bibitem[\protect\citeauthoryear{Edman et al.}{2012}]{edman2012} Edman J.~P., Barton E.~J., Bullock J.~S., 2012, MNRAS, 424, 1454 

\bibitem[\protect\citeauthoryear{Ellison et al.}{2008}]{ellison2008} Ellison S.~L., Patton D.~R., Simard L., McConnachie A.~W., 2008, AJ, 135, 1877

\bibitem[\protect\citeauthoryear{Ellison et al.}{2010}]{ellison2010} Ellison S.~L., Patton D.~R., Simard L., McConnachie A. W., Baldry I. K., Mendel J. T., 2010, MNRAS, 407, 1514 

\bibitem[\protect\citeauthoryear{Ellison et al.}{2011}]{ellison2011} Ellison S.~L., Patton D.~R., Mendel J.~T., Scudder J.~M., 2011, MNRAS, 418, 2043 

\bibitem[\protect\citeauthoryear{Faber et al.}{2007}]{faber2007} Faber S.~M. et al., 2007, ApJ, 665, 265

\bibitem[\protect\citeauthoryear{Farouki \& Shapiro}{1981}]{farouki1981} Farouki R., Shapiro S.~L., 1981, ApJ, 243, 32 



\bibitem[\protect\citeauthoryear{Font et al.}{2008}]{font2008} Font A.~S. et al., 2008, MNRAS, 389, 1619 

\bibitem[\protect\citeauthoryear{Fujita}{1998}]{fujita1998} Fujita Y., 1998, ApJ, 509, 

\bibitem[\protect\citeauthoryear{Gerke et al.}{2005}]{gerke2005} Gerke B.~F. et al., 2005, ApJ, 625, 6 

\bibitem[\protect\citeauthoryear{Gunn \& Gott}{1972}]{gunn1972} Gunn J.~E., Gott J.~R.~I., 1972, ApJ, 176, 1 




\bibitem[\protect\citeauthoryear{Hubble}{1926}]{hubble1926} Hubble E.~P., 1926, ApJ, 64, 321

\bibitem[Iovino et al.(2010)]{iovino2010} Iovino A. et al., 2010, A\&A, 509, A40 

\bibitem[\protect\citeauthoryear{Kang \& van den Bosch}{2008}]{kang2008} Kang X., van den Bosch F.~C., 2008, ApJL, 676, L101

\bibitem[\protect\citeauthoryear{Kauffmann et al.}{1993}]{kauffmann1993} Kauffmann G., White S.~D.~M., Guiderdoni B., 1993, MNRAS, 264, 201 

\bibitem[\protect\citeauthoryear{Kauffmann et al.}{2003}]{kauffmann2003} Kauffmann G. et al., 2003, MNRAS, 341, 33 


\bibitem[\protect\citeauthoryear{Kere{\v s} et al.}{2009}]{keres2009} Kere{\v s} D., Katz N., Fardal M., Dav{\'e} R., Weinberg D.~H., 2009, MNRAS, 395, 160 

\bibitem[\protect\citeauthoryear{Kimm et al.}{2009}]{kimm2009} Kimm T. et al., 2009, MNRAS, 394, 1131

\bibitem[\protect\citeauthoryear{Klypin et al.}{1999}]{klypin1999} Klypin A., Gottl{\"o}ber S., Kravtsov A.~V., Khokhlov A.~M., 1999, ApJ, 516, 530 

\bibitem[Knobel et al.(2013)]{knobel2013} Knobel C. et al., 2013, ApJ, 769, 24 

\bibitem[\protect\citeauthoryear{Kravtsov et al.}{1997}]{kravtsov1997} Kravtsov A.~V., Klypin A.~A., Khokhlov A.~M.\ 1997, ApJS, 111, 73 

\bibitem[\protect\citeauthoryear{Kravtsov et al.}{2004}]{kravtsov2004} Kravtsov A.~V., Berlind A.~A., Wechsler R.~H., Klypin A.~A., Gottl{\"o}ber S., Allgood B., Primack J.~R., 2004, ApJ, 609, 35

\bibitem[\protect\citeauthoryear{Larson et al.}{1980}]{larson1980} Larson R.~B., Tinsley B.~M., Caldwell C.~N., 1980, ApJ, 237, 692 

\bibitem[\protect\citeauthoryear{Makino \& Hut}{1997}]{makino1997} Makino J., Hut P., 1997, ApJ, 481, 83 


\bibitem[\protect\citeauthoryear{McGee et al.}{2009}]{mcgee2009} McGee S.~L., Balogh M.~L., Bower R.~G., Font A.~S., McCarthy I.~G., 2009, MNRAS, 400, 937

\bibitem[\protect\citeauthoryear{McGee et al.}{2011}]{mcgee2011} McGee S.~L., Balogh M.~L., Wilman D.~J., Bower R. G., Mulchaey J. S., Parker L. C., Oemler A., Jr., 2011, MNRAS, 413, 996 

\bibitem[\protect\citeauthoryear{Mihos \& Hernquist}{1996}]{mihos1996} Mihos J.~C., Hernquist L., 1996, ApJ, 464, 641 


\bibitem[\protect\citeauthoryear{Moore et al.}{1996}]{moore1996} Moore B., Katz N., Lake G., Dressler A., Oemler A., 1996, Nat, 379, 613

\bibitem[\protect\citeauthoryear{Nichols \& Bland-Hawthorn}{2011}]{nichols2011} Nichols M., Bland-Hawthorn J., 2011, ApJ, 732, 17 

\bibitem[\protect\citeauthoryear{Pasquali et al.}{2010}]{pasquali2010} Pasquali A., Gallazzi A., Fontanot F., van den Bosch F.~C., De Lucia G., Mo H., Yang X., 2010, MNRAS, 407, 937

\bibitem[\protect\citeauthoryear{Patton et al.}{2011}]{patton2011} Patton D.~R., Ellison S.~L., Simard L., McConnachie A.~W., Mendel J.~T., 2011, MNRAS, 412, 591 

\bibitem[Phillips et al.(2013)]{phillips2013} Phillips J.~I., Wheeler C., Boylan-Kolchin M., Bullock J. S., Cooper, M. C., Tollerud, E. J., 2013, arXiv:1307.3552 

\bibitem[\protect\citeauthoryear{Postman \& Geller}{1984}]{postman1984} Postman M., Geller M.~J., 1984, ApJ, 281, 95 

 
\bibitem[\protect\citeauthoryear{Petrosian}{1976}]{petrosian1976} Petrosian V., 1976, ApJL, 209, L1 

\bibitem[\protect\citeauthoryear{Quilis et al.}{2000}]{quilis2000} Quilis V., Moore B., Bower R., 2000, Science, 288, 1617

\bibitem[\protect\citeauthoryear{Rines et al.}{2005}]{rines2005} Rines K., Geller M.~J., Kurtz M.~J., Diaferio A., 2005, AJ, 130, 1482


\bibitem[\protect\citeauthoryear{Schlegel et al.}{1998}]{schlegel1998} Schlegel D.~J., Finkbeiner D.~P., Davis M., 1998, ApJ, 500, 525

\bibitem[\protect\citeauthoryear{Scudder et al.}{2012}]{scudder2012} Scudder J.~M., Ellison S.~L., Torrey P., Patton D.~R., Mendel J.~T., 2012, MNRAS, 426, 549 


\bibitem[\protect\citeauthoryear{Skibba}{2009}]{skibba2009} Skibba R.~A., 2009, MNRAS, 399, 966

\bibitem[\protect\citeauthoryear{Sol Alonso et al.}{2006}]{solalonso2006} Sol Alonso M., Lambas D.~G., Tissera P., Coldwell G., 2006, MNRAS, 367, 1029 

\bibitem[\protect\citeauthoryear{Somerville \& Kolatt}{1999}]{somerville1999} Somerville R.~S., Kolatt T.~S., 1999, MNRAS, 305, 1 


\bibitem[\protect\citeauthoryear{Stewart et al.}{2008}]{stewart2008} Stewart K.~R., Bullock J.~S., Wechsler R.~H., Maller A.~H., Zentner A.~R., 2008, ApJ, 683, 597 

\bibitem[\protect\citeauthoryear{Strateva et al.}{2001}]{strateva2001} Strateva I. et al., 2001, AJ, 122, 1861

\bibitem[\protect\citeauthoryear{Strauss et al.}{2002}]{strauss2002} Strauss M.~A. et al., 2002, AJ, 124, 1810


\bibitem[\protect\citeauthoryear{Tanaka et al.}{2005}]{tanaka2005} Tanaka M., Kodama T., Arimoto N., Okamura S., Umetsu K., Shimasaku K., Tanaka I., Yamada T., 2005, MNRAS, 362, 268 

\bibitem[\protect\citeauthoryear{Tollerud et al.}{2011}]{tollerud2011} Tollerud E.~J., Boylan-Kolchin M., Barton E.~J., Bullock J.~S., Trinh C.~Q., 2011, ApJ, 738, 102 

\bibitem[\protect\citeauthoryear{Toomre \& Toomre}{1972}]{toomre1972} Toomre A., Toomre J., 1972, ApJ, 178, 623

\bibitem[\protect\citeauthoryear{van den Bosch et al.}{2008}]{vdb2008} van den Bosch F.~C., Aquino D., Yang X., Mo H.~J., Pasquali A., McIntosh D.~H., Weinmann S.~M., Kang X., 2008, MNRAS, 387, 79 

\bibitem[\protect\citeauthoryear{von der Linden et al.}{2007}]{vonderlinden2007} von der Linden A., Best P.~N., Kauffmann G., White S.~D.~M., 2007, MNRAS, 379, 867 

\bibitem[\protect\citeauthoryear{von der Linden et al.}{2010}]{vonderlinden2010} von der Linden A., Wild V., Kauffmann G., White S.~D.~M., Weinmann S., 2010, MNRAS, 404, 1231

\bibitem[\protect\citeauthoryear{Wechsler et al.}{2006}]{wechsler2006} Wechsler R.~H., Zentner A.~R., Bullock J.~S., Kravtsov A.~V., Allgood B., 2006, ApJ, 652, 71

\bibitem[\protect\citeauthoryear{Weinmann et al.}{2006a}]{weinmann2006a} Weinmann S.~M., van den Bosch F.~C., Yang X., Mo H.~J., 2006, MNRAS, 366, 2

\bibitem[\protect\citeauthoryear{Weinmann et al.}{2006b}]{weinmann2006b} Weinmann S.~M., van den Bosch F.~C., Yang X., Mo H.~J., Croton D.~J., Moore B., 2006, MNRAS, 372, 1161 

\bibitem[\protect\citeauthoryear{Weinmann et al.}{2009}]{weinmann2009} Weinmann S.~M., Kauffmann G., van den Bosch F.~C., Pasquali A., McIntosh D.~H., Mo H., Yang X., Guo Y., 2009, MNRAS, 394, 1213

\bibitem[\protect\citeauthoryear{Weinmann et al.}{2010}]{weinmann2010} Weinmann S.~M., Kauffmann G., von der Linden A., De Lucia G., 2010, MNRAS, 406, 2249
 
\bibitem[\protect\citeauthoryear{Wetzel et al.}{2012}]{wetzel2012} Wetzel A.~R., Tinker J.~L., Conroy C., 2012, MNRAS, 424, 232

\bibitem[\protect\citeauthoryear{White \& Rees}{1978}]{white1978} White S.~D.~M., Rees M.~J., 1978, MNRAS, 183, 341

\bibitem[\protect\citeauthoryear{Zentner \& Bullock}{2003}]{zentner2003} Zentner A.~R., Bullock J.~S., 2003, ApJ, 598, 49

\bibitem[\protect\citeauthoryear{Zentner et al.}{2005}] {zentner2005} Zentner A.~R., Berlind A.~A., Bullock J.~S., Kravtsov A.~V., Wechsler R.~H., 2005, ApJ, 624, 505

\bibitem[\protect\citeauthoryear{Yang et al.}{2005}]{yang2005} Yang X., Mo H.~J., van den Bosch F.~C., Jing Y.~P., 2005, MNRAS, 356, 1293 


\bibitem[\protect\citeauthoryear{York et al.}{2000}]{york2000} York D.~G. et al., 2000, AJ, 120, 1579

\end{thebibliography}
\end{document}